\documentclass[fleqn,usenatbib]{mnras}

\usepackage{newtxtext,newtxmath}
\usepackage[T1]{fontenc}
\usepackage{graphicx}	% Including figure files
\usepackage{amsmath}	% Advanced maths commands
\usepackage{gensymb}

\newcommand{\hi}{H\textsc{i}\ }
\newcommand{\hinospace}{\textrm{H\textsc{i}}}
\newcommand{\secref}[1]{\hyperref[#1]{Section~\ref*{#1}}}

\title[eBOSS - GBT H\textsc{i} intensity mapping  cross-correlations]{H\textsc{i} constraints from the cross-correlation of eBOSS galaxies and Green Bank Telescope intensity maps
}

\author[L. Wolz et al.]{Laura Wolz$^{1}$\thanks{E-mail:laura.wolz@manchester.ac.uk},
Alkistis Pourtsidou$^{2}$,
Kiyoshi W. Masui$^{3,4}$, Tzu-Ching Chang$^{5,6,7}$,\newauthor Julian E. Bautista$^{8,9}$, Eva-Maria M\"uller$^{10}$, Santiago Avila$^{11,12}$, David Bacon$^{9}$,\newauthor  Will J. Percival$^{13,14,15}$, Steven Cunnington$^{2}$, Chris Anderson$^{16}$, Xuelei Chen$^{17}$,  \newauthor  Jean-Paul Kneib$^{18}$, Yi-Chao Li$^{19}$, Yu-Wei Liao$^{7}$, Ue-Li Pen$^{20}$, Jeffrey B.  Peterson$^{21}$,\newauthor Graziano Rossi$^{22}$, Donald P. Schneider$^{23,24}$, Jaswant Yadav$^{25}$, Gong-Bo Zhao$^{17,9}$
\\
$^{1}$Jodrell Bank Centre for Astrophysics, Department of Physics and Astronomy, The University of Manchester, Manchester M13 9PL, UK\\
$^{2}$School of Physics and Astronomy, Queen Mary University of London, Mile End Road, London E1 4NS, UK\\
$^{3}$MIT Kavli Institute for Astrophysics and Space Research, Massachusetts Institute of Technology, 77 Massachusetts Ave, Cambridge, MA 02139, USA\\
$^{4}$Department of Physics, Massachusetts Institute of Technology, 77 Massachusetts Ave, Cambridge, MA 02139, USA\\
$^{5}$Jet Propulsion Laboratory, California Institute of Technology, Pasadena, CA 91101, USA\\
$^{6}$California Institute of Technology, Pasadena, CA 91125, USA\\
$^{7}$Academia Sinica Institute of Astronomy and Astrophysics, Roosevelt Rd, Taipei 10617, Taiwan\\
$^{8}$Aix Marseille Univ, CNRS/IN2P3, CPPM, Marseille, France\\
$^{9}$Institute of Cosmology \& Gravitation, University of Portsmouth, Dennis Sciama Building, Portsmouth, PO1 3FX, United Kingdom\\
$^{10}$Department of Physics, University of Oxford, Denys Wilkinson Building, Keble Road, Oxford OX1 3RH\\
$^{11}$Departamento de F\'isica Te\'orica, Facultad de Ciencias, Universidad Aut\'onoma de Madrid, 28049 Cantoblanco, Madrid, Spain\\
$^{12}$Instituto de F\'isica Teorica UAM-CSIC, Universidad Aut\'onoma de Madrid, 28049 Cantoblanco, Madrid, Spain \\
$^{13}$Waterloo Centre for Astrophysics, University of Waterloo, Waterloo, ON N2L 3G1, Canada\\
$^{14}$Department of Physics and Astronomy, University of Waterloo, Waterloo, ON N2L 3G1, Canada\\
$^{15}$Perimeter Institute for Theoretical Physics, 31 Caroline St. North, Waterloo, ON N2L 2Y5, Canada\\
$^{16}$Department of Physics, University of Wisconsin Madison, 1150 University Ave, Madison WI 53703, USA\\
$^{17}$National Astronomical Observatories, Chinese Academy of Sciences,
Beijing 100101, China\\
$^{18}$Institute of Physics, Laboratory of Astrophysics, Ecole Polytechnique Federale de Lausanne
(EPFL), Observatoire de Sauverny, 1290 Versoix, Switzerland\\
$^{19}$Department of Physics \& Astronomy, University of the Western Cape, Cape Town 7535, South Africa\\
$^{20}$
Canadian Institute for Theoretical Astrophysics, University of Toronto, 60 St. George St., Toronto Ontario, M5S 3H8, Canada\\
$^{21}$Department of Physics, Carnegie Mellon University. Pittsburgh. PA, USA\\
$^{22}$Department of Astronomy and Space Science, Sejong University, 209, Neungdong-ro, Gwangjin-gu, Seoul, South Korea\\
$^{23}$Department of Astronomy and Astrophysics, The Pennsylvania State University, University Park, PA 16802\\
$^{24}$
Institute for Gravitation and the Cosmos, The Pennsylvania State University, University
Park, PA 16802, USA\\
$^{25}$Central University of Haryana, Jant-Pali, Mahendergarh - 123031, India
}

\date{Accepted XXX. Received YYY; in original form ZZZ}

\pubyear{2020}

\begin{document}
\label{firstpage}
\pagerange{\pageref{firstpage}--\pageref{lastpage}}
\maketitle

\begin{abstract}
We present the joint analysis of Neutral Hydrogen (H\textsc{i}) Intensity Mapping observations with three galaxy samples: the Luminous Red Galaxy (LRG) and Emission Line Galaxy (ELG) samples from the eBOSS survey, and the WiggleZ Dark Energy Survey sample. The H\textsc{i} intensity maps are Green Bank Telescope observations of the redshifted $21\rm cm$ emission on $100 \, {\rm deg}^2$ covering the redshift range $0.6<z<1.0$. We process the data by separating and removing the foregrounds present in the radio frequencies with \textsc{FastICA}. We verify the quality of the foreground separation with mock realisations, and construct a transfer function to correct for the effects of foreground removal on the H\textsc{i} signal. We cross-correlate the cleaned H\textsc{i} data with the galaxy samples and study the overall amplitude as well as the scale-dependence of the power spectrum. We also qualitatively compare our findings with the predictions by a semi-analytic galaxy evolution simulation. 
The cross-correlations constrain the quantity $\Omega_{\textrm{H\textsc{i}}} b_{\textrm{H\textsc{i}}} r_{\textrm{H\textsc{i}},{\rm opt}}$ at an effective scale $k_{\rm eff}$, where $\Omega_\textrm{H\textsc{i}}$ is the H\textsc{i} density fraction, $b_\textrm{H\textsc{i}}$ is the H\textsc{i} bias, and $r_{\textrm{H\textsc{i}},{\rm opt}}$ the galaxy-hydrogen correlation coefficient, which is dependent on the H\textsc{i} content of the optical galaxy sample. 
At $k_{\rm eff}=0.31 \, h/{\rm Mpc}$
we find $\Omega_{\textrm{H\textsc{i}}} b_{\textrm{H\textsc{i}}} r_{\textrm{H\textsc{i}},{\rm Wig}} = [0.58 \pm 0.09 \, {\rm (stat) \pm 0.05 \, {\rm (sys)}}] \times 10^{-3}$ for GBT-WiggleZ, $\Omega_{\textrm{H\textsc{i}}} b_{\textrm{H\textsc{i}}} r_{\textrm{H\textsc{i}},{\rm ELG}} = [0.40 \pm 0.09 \, {\rm (stat) \pm 0.04 \, {\rm (sys)}}] \times 10^{-3}$ for GBT-ELG, and $\Omega_{\textrm{H\textsc{i}}} b_{\textrm{H\textsc{i}}} r_{\textrm{H\textsc{i}},{\rm LRG}} = [0.35 \pm 0.08 \, {\rm (stat) \pm 0.03 \, {\rm (sys)}}] \times 10^{-3}$ for GBT-LRG, at $z\simeq 0.8$. We also report results at $k_{\rm eff}=0.24 \, h/{\rm Mpc}$ and $k_{\rm eff}=0.48 \, h/{\rm Mpc}$.
With little information on H\textsc{i} parameters beyond our local Universe, these are amongst the most precise constraints on neutral hydrogen density fluctuations in an underexplored redshift range.
\end{abstract}


%%%%%%%%%%%%%%%% BODY OF PAPER %%%%%%%%%%%%%%%%%%

\section{Introduction}

The redshifted 21cm emission from Neutral Hydrogen (\hinospace) gas provides an alternative view into the structure, dynamics, and evolution of galaxies. \hi gas is the fundamental fuel for molecular gas and star formation and plays an essential role in galaxy formation and evolution and models thereof. 
Blind \hi surveys of the local Universe provide constraints on the \hi abundance via the \hi mass function  \citep{Jones:2020ik, 2003AJ....125.2842Z} and the global \hi abundance $\Omega_{\hinospace}=(4.3\pm0.3)10^{-4}H_0/70$ \citep{Martin:2010ij}. Spectral stacking techniques have also been used (see e.g. \citealt{Hu:2019xmd} and references therein). 

Targeted deep surveys investigate the \hi scaling relations with galaxy properties such as stellar mass, star formation activity, or star formation efficiency with multi-wavelength data. It has been inferred that cold gas properties are tightly related to their star-forming properties and less to their morphology, with scatter on the relations being driven by inflows mechanisms and dynamics \citep{2019MNRAS.490.4060C, Chen:2019gz}. \hi gas mass has been found to strongly anti-correlate with stellar mass, particularly when traced by NUV-r colour \citep{2018MNRAS.476..875C}. Multiple studies on the \hi deficiency in high density regions such as the VIRGO cluster confirm the high impact of environment on atomic gas abundance (see \citealt{Cortese:2011sj, 2014MNRAS.444..667D, Reynolds:2020tm}). \citet{2020arXiv200914585B} studied environmental effects using an infra-red selected sample of \hi detections finding a reduced scatter in scaling relations for isolated galaxies. Some investigations have been made into the relation between \hi and its host halo mass to constrain a \hi halo occupation distribution, see for example \citet{Guo:2020kt} or \citet{Paul:2017bi}. The most important limitations of all blind and targeted \hi surveys are their sensitivity limitations on relatively \hinospace-rich galaxy samples, as well as volume-limited sample sizes. Additionally, there is little information on \hi abundances and scaling relations beyond our local Universe \citep{Crighton:2015pza, Padmanabhan:2015wja,Hu:2019xmd}.

The technique of \hi intensity mapping has been proposed to perform fast observations of very large cosmic volumes in a wide redshift range. Intensity mapping does not rely on detecting individual galaxies, but instead measures the integrated redshifted spectral line emission without sensitivity cuts in large voxels on the sky, whith the voxel volume determined by the radio telescope beam and frequency channelisation, see e.g. \citep{Battye2004, Chang_2008, Wyithe_2009, Mao_2008, peterson2009, Chang:2010jp, Seo_2010, Ansari_2012}. Using the \hi signal as a biased tracer for the underlying matter distribution, it is possible to probe the large-scale structure of the Universe, and constrain both, global \hi properties and cosmological parameters. Particularly, the amplitude of the \hi intensity mapping clustering signal scales with the global \hi energy density $\Omega_{\rm HI}$ and can constrain it for various redshifts. 

The next few years will see data from a number of \hi intensity mapping experiments, for example the proposed MeerKLASS survey at the Square Kilometre Array (SKA) precursor MeerKAT \citep{Santos:2017qgq,Wang:2020lkn}, an \hi survey at the 500m dish telescope FAST \citep{Hu:2019okh}, and multiple surveys with the SKA using the single-dish mode of operation \citep{Battye:2012tg, Bull:2014rha, Santos:2017qgq,Bacon:2018dui}. Other international experiments include the CHIME project \citep{Bandura:2014gwa}, HIRAX \citep{Newburgh:2016mwi}, and Tianlai \citep{Li:2020ast, Wu:2020jwm}. 

The observed intensity maps suffer from foreground contamination from Galactic and extra-galactic sources. Our own Galaxy emits high synchroton and free-free emission up to three orders of magnitude brighter than the redshifted 21cm line \citep{Di_Matteo_2002}, which need to be subtracted from the data (see e.g. \citealt{Wolz:2013wna,Alonso:2014dhk, Shaw:2014khi, Olivari:2015dc,Cunnington:2019lvb, Carucci:2020ca}). 
To-date, the intensity mapping signal has not been detected in auto-correlation due to calibration errors, radio frequency interference, residual foregrounds and noise systematics \citep{Switzer_2013,Switzer_2015, 2018MNRAS.478.2416H, 2020arXiv200701767L}. 
The impact of the contaminations can be reduced by cross-correlating the \hi signal with optical surveys. The first successful detection with Green Bank Telescope (GBT) data has been achieved at $0.6<z<1.0$ using the cross-correlations with  the DEEP2 survey \citep{Chang:2010jp}, followed by the cross-correlations with the WiggleZ Dark Energy survey \citep{Masui:2012zc}. The GBT-WiggleZ correlations at $z=0.8$ have constrained the combination of the \hi abundance $\Omega_\hinospace$ and linear \hi bias $b_\hinospace$, finding $\Omega_\hinospace b_\hinospace r_{\hinospace,{\rm Wig}} = [4.3 \pm 1.1] \times 10^{-4}$, where $r_{\hinospace,{\rm Wig}}$ is the galaxy-\hi cross-correlation coefficient. The significance of detection was $7.4\sigma$ for the combined 1hr and 15hr fields observations \citep{Masui:2012zc}. 

More recently, the Parkes radio telescope reported a cross-correlation detection at $z\simeq 0.1$ using galaxies from the 2dF survey \citep{Anderson_2018}. In this study, upon dividing the galaxies into red and blue colours, a drop in amplitude on small scales was detected for the red  sample. This result is in agreement with aforementioned studies on \hi in dense environments as well as with theoretical predictions on the \hi-galaxy cross-correlation of a correlation coefficient dependent on the \hi content of the optical galaxy sample \citep{2016MNRAS.458.3399W}. Additionally, it is also predicted that the amplitude of the shot noise on the cross-power spectra scales with the averaged \hi mass of the galaxy sample \citep{2017MNRAS.470.3220W}. 

In this work, we present the analysis of the extended and deepened 1-hr field observations from the previous study in \citet{Masui:2012zc}. We apply the foreground subtraction technique \textsc{FastICA} as outlined in \citet{Wolz:2015lwa} and, for the first time, construct the \textsc{FastICA} transfer function using mock lognormal simulations. We cross-correlate the \hi intensity mapping data with three distinct galaxy samples,
the Emission Line Galaxy (ELG) and Luminous Red Galaxy (LRG) samples from the eBOSS survey \citep{Raichoor:2020jcl,Ross:2020lqz,Alam:2020sor} as well as the previously considered WiggleZ survey \citep{blake2011}. This leads to a robust confirmation of detection with multiple galaxy samples, as well as a first attempt to quantify the cross-correlation coefficient between \hi and the galaxy sample properties. We also qualitatively compare our measurements with predictions from the semi-analytic galaxy evolution model DARK SAGE \citep{2016MNRAS.461..859S} to investigate the \hi contents of the samples. Finally, we use the cross-correlation measurements to constrain the quantity $\Omega_\hinospace b_\hinospace r_{\hinospace,{\rm opt}}$, and also provide estimates for $\Omega_\hinospace$ using external estimates for $b_\hinospace$ and $r_{\hinospace,{\rm opt}}$.

The paper is organised as follows: In \secref{sec:data}, we describe the GBT intensity maps, and the WiggleZ and eBOSS galaxy samples. We also give a brief description of our simulations. In \secref{sec:FGremoval}, we outline the application of the \textsc{FastICA} technique to the GBT maps, as well as the construction of the foreground transfer function. In \secref{sec:PSresults} we present and discuss our cross-correlation results. In \secref{sec:HIconstraints} we derive the \hi constraints. We conclude in \secref{sec:conclusions}. The appendix contains details on our mock galaxy selection in \autoref{app:samples} as well as figures of our covariance analysis in \autoref{appb}.

\section{Description of data products}
\label{sec:data}

\subsection{Green Bank Telescope intensity maps}
The \hi intensity mapping data from the Green Bank Telescope (GBT) used in this study is located in the 1hr field of the WiggleZ Dark Energy survey at right ascension $5.43\degree < {\rm RA} <18.9 \degree$ and declination $-2.55\degree < {\rm DEC} < 4.8\degree$. This field was observed with the receiver band at $700<\nu<900 \, \rm{MHz}$, which results in a 21cm redshift range of $0.6<z<1.0$. The data is divided into $N_{\nu}=256$ frequency channels with width $\delta\nu=0.78 \, \rm MHz$, after rebinning from the original 2048 correlator channels. The observational spatial resolution of the maps, quantified by the full width half maximum (FWHM) of the GBT telescope beam, evolves from $\rm{FWHM}\approx0.31\deg$ at $\nu=700 \, \rm{MHz}$ to $\rm{FWHM}\approx 0.25\deg$ at $\nu=900 \, \rm{MHz}$. The maps are pixelised with spatial resolution angle of $\delta\theta\approx \delta\phi=0.067\deg$, which results in $N_{\rm RA}=217$ pixels in right ascension and $N_{\rm DEC}=119$ pixels in declination. The pixel size was chosen such that approximately 4 pixels cover the beam at mid-frequency $\nu\approx800 \, \rm MHz$, and the instrumental noise can be approximated as uncorrelated between pixels.
The maps are an extended version of the previously published observations described in \citet{Masui:2012zc} with added scans to increase the area to $100\deg^2$ and survey depth to $100 \, \rm hrs$ total integration time collected from 2010-2015. The details on Radio Frequency Interference (RFI) flagging, calibration, and map making procedures can be found in \citet{Masui:2012zc, Switzer_2013, 2013PhDT.......570M}.

\begin{figure*}
    \centering
    \includegraphics[width=0.8\textwidth]{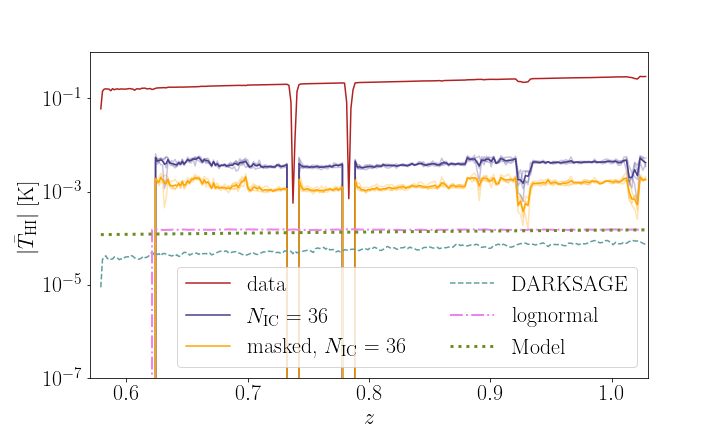}
     \caption{Mean of the absolute temperature of the GBT intensity maps as a function of redshift, binned into 256 frequency channels. The solid lines represent the mean over the 4 GBT seasons with original data (red), the \textsc{FastICA} foreground subtracted data with $N_{\rm IC}=36$ (purple), and the masked, \textsc{FastICA} foreground subtracted data with $N_{\rm IC}=36$ (yellow). The faded purple and yellow lines indicate the individual seasons. The green dotted line represents the analytical brightness temperature prediction from \autoref{eq:thi}, the pink dashed line the averaged temperature of the lognormal simulations used for the foreground removal transfer function (see \secref{sec:FGremoval} for details), and the teal dashed line the numerical prediction from the DARK SAGE simulation described in \secref{sec:data}.}
    \label{fig:Tmean_redshift}
\end{figure*}

As described in previous studies, the GBT intensity maps suffer a number of instrumental systematic effects. To reduce the impact of the systematic effects, the following measures have been taken:
\begin{itemize}
    \item RFI and resonance: The data is contaminated by RFI and two telescope resonance frequencies. \autoref{fig:Tmean_redshift} shows the mean absolute temperature of each channel as a function of redshift. The red line shows the initial data with strong RFI contamination at the lowest redshift as well as towards the highest redshift end. The RFI flagging causes an overall signal loss of $\approx 11\%$, more details on the RFI flagging process can be found in \citep{Switzer_2013}. The two telescope resonances can be seen at $\nu=798 \, \rm MHz$ and $\nu=817 \, \rm MHz$ which corresponds to the dips in amplitude seen at $z=0.78$ and $z=0.74$. To minimise these effects, we dismiss the lowest 30 channels in redshift and the intervals around the resonances before the foreground removal.
    \item Sub-seasons: The time-ordered data is divided into 4 seasons $\rm \{A, B, C, D\}$. Thermal noise is uncorrelated between these seasons, which have been chosen to have similar integration depth and coverage \citep{Switzer_2013}. More specifically, the Gaussian sampling noise and time-dependent RFI in each season are independent, however, observational systematics in seasons can correlate. The individual season data is shown as faded purple and yellow lines in \autoref{fig:Tmean_redshift}.
    \item Masking: The noise properties are highly anisotropic towards the spatial edges of the map due to the scanning strategy and resulting anisotropic survey depth. We therefore mask out 15 pixels per side, which significantly reduces residual anisotropic noise in the foreground subtracted maps. About an order of magnitude decrease of the mean temperature of the maps is found comparing the original and masked foreground subtracted data marked by the purple and yellow lines in \autoref{fig:Tmean_redshift}. 
    The solid purple and yellow lines show the signal averaged over the four seasons, and the faded lines around them show the individual seasons.
    \item Beam: The beam of the instrument can be approximated by a symmetric Gaussian function with a frequency-dependent FWHM with maximum ${\rm FWHM}_{\rm max}\approx 0.31\deg$. In order to aid the data analysis as well as to minimise systematics caused by polarisation leakage of the receiver \citep{Switzer_2013}, we convolve the data to a common Gaussian beam with ${\rm FWHM}=1.4\,{\rm FWHM_{max}}$, which results in an angular resolution of ${\rm FWHM}=0.44\deg$. This strategy is adopted as polarization leakage is considered the most significant contaminant in the data. However, we acknowledge that this would not be an optimal strategy to mitigate effects of beam chromaticity, as shown in \cite{FgSKA}.
\end{itemize}
\autoref{fig:Tmean_redshift} shows that even after applying these measures and removing foregrounds modelled by 36 Independent Components, the mean temperature of the \hi maps is about an order of magnitude higher than the theoretically predicted \hi brightness temperature. We model this following \citet{Chang:2010jp} and \citet{Masui:2012zc} as:
\begin{equation}
    T_{\rm HI}(z)=0.29\frac{\Omega_{\rm HI}}{10^{-3}} \left(\frac{\Omega_m+\Omega_{\Lambda}(1+z)^{-3}}{0.37}\right)^{-0.5}\left( \frac{1+z}{1.8}\right)^{0.5} {\rm mK} \,
    \label{eq:thi}
\end{equation}
which is shown as the green dotted line. We are unable to directly detect the \hi signal with our current pipeline in this systematics dominated data.
\subsection{Galaxy samples}
In this study, we consider three galaxy samples overlapping with the \hi intensity maps in the 1hr field. We use the WiggleZ Dark Energy Survey galaxy sample based on \citet{blake2011} as previously presented in \citet{Masui:2012zc}. And, for the first time, we use the SDSS Emission Line Galaxy (ELG) and Luminous Red Galaxy (LRG) sample of the eBOSS survey (DR16) for the \hinospace-galaxy cross-correlation analysis. 

\begin{figure}
    \centering
    \includegraphics[width=\columnwidth]{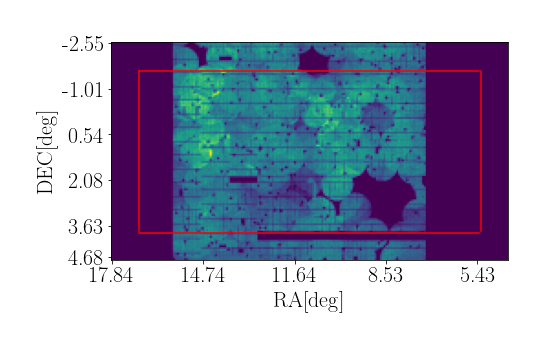} \\
     \includegraphics[width=\columnwidth]{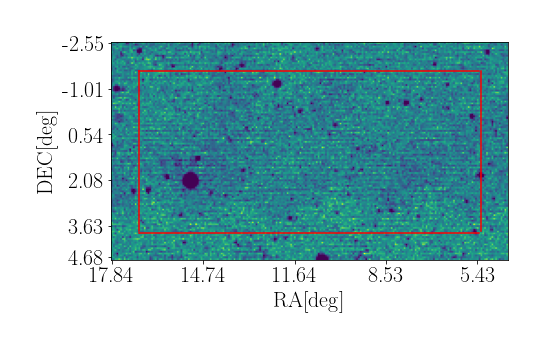}
 \includegraphics[width=\columnwidth]{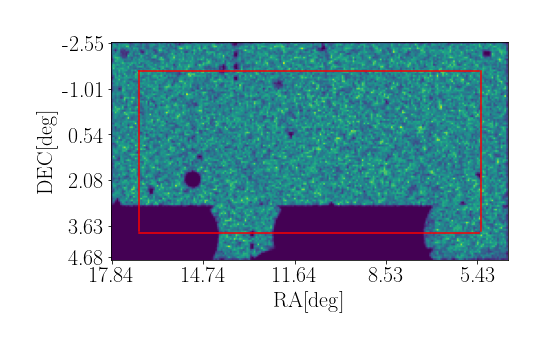}

    \caption{Spatial footprint of the galaxy samples. \emph{From top to bottom:} WiggleZ, ELG, and LRG samples. The survey window is binned on the same spatial pixelisation as the GBT data with pixel size of $\delta\theta=\delta\phi=0.067\deg$.
    }
    \label{fig:Sel_gal}
\end{figure}

In \autoref{fig:Sel_gal}, we show the spatial footprint of each survey in the 1hr field, where dark patches indicate unobserved regions and the red lines mark the edge masking as part of the systematics mitigation of the GBT data. The LRG and WiggleZ samples both have a reduced spatial overlap with the GBT data as it has unobserved regions, however, since we introduce the red mask, this effect is somewhat diminished. The ELG sample has the most complete overlap with the GBT data.

{\bf WiggleZ} -
The WiggleZ galaxies are part of the WiggleZ Dark Energy Survey \citep{Drinkwater}, a large-scale spectroscopic survey of emission-line galaxies selected from UV and optical imaging. These are active, highly star-forming objects, and it has been suggested that they contain a large amount of \hi gas to fuel their star-formation.
The selection function \citep{2010MNRAS.406..803B} has angular dependence determined primarily by the UV selection, and redshift coverage favouring the $z = 0.6$ end of the radio band. The galaxies are binned into volumes with the same pixelization as the radio maps and divided by the selection function, and we consider the cross-power with respect to optical over-density. 

{\bf eBOSS ELG} - The extended Baryon Oscillation Spectroscopic Survey (eBOSS; \citealt{dawson_sdss-iv_2016}), is part of the SDSS-IV experiment \citep{2017AJ....154...28B}, and has spectroscopically observed $173,736$ ELGs in the redshift range $0.6<z<1.1$ \citep{Raichoor:2020jcl}. Targets were colour-selected from the DECaLS photometric survey, with an algorithm designed to select OII emitting galaxies with high star-formation rates. Spectra were then obtained using the BOSS spectrographs \citep{2013AJ....146...32S} mounted on the 2.5-meter Sloan telescope \citep{2006AJ....131.2332G}. Details of the sample, including standard Baryon Acoustic Oscillation (BAO) and Redshift Space Distortion (RSD) measurements can be found in \citet{Raichoor:2020jcl,Tamone:2020qrl,deMattia:2020fkb}.

{\bf eBOSS LRG} - Luminous Red Galaxies were observed by eBOSS from a target sample selected \citep{Prakash:2015eua} from SDSS DR13 photometric data \citep{Albareti:2016xlm}, combined with infrared observations from the WISE satellite \citep{2016AJ....151...36L}. This sample was selected to be composed of large, old, strongly-biased galaxies, typically found in high mass haloes. In total, the sample contains $174,816$ LRGs with measured redshifts between $0.6<z<1.0$. In our analysis we do not combine the eBOSS LRGs with the $z > 0.6$ BOSS CMASS galaxies as in the standard BAO and RSD measurements \citep{Bautista:2020ahg,Gil-Marin:2020bct}. Possible systematics related to the eBOSS LRG sample have been quantified via realistic N-body-based mocks in \citep{2021MNRAS.505..377R}. The cosmological interpretation of the BAO and RSD results from all eBOSS samples was presented in \citet{Alam:2020sor}.

\begin{figure}
    \centering
    \includegraphics[width=\columnwidth]{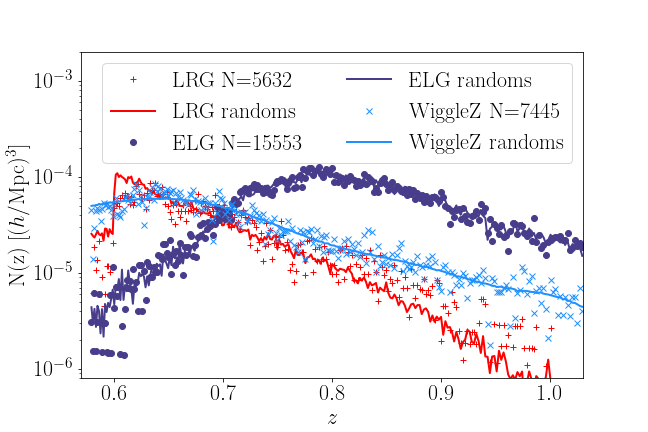}
    \caption{Galaxy density distribution with redshift. The solid lines represent the mean of the random catalogues used to determine the selection function, and the markers show the data points of the samples. }
    \label{fig:Nz_gal}
\end{figure}

\autoref{fig:Nz_gal} shows the galaxy density distribution with redshift, $N(z)$, where we binned the data according to the frequency bins of the GBT intensity mapping data. This implies that the bin size is constant in frequency rather than redshift, and the co-moving volume of the bins evolves with redshift. The line-of-sight resolution is very high with an average redshift bin size of $\delta z \approx 0.0016$. The galaxy density normalisation has taken into account the evolving co-moving volume of the bins. 
We can see that both the WiggleZ galaxy and eBOSS LRG samples peak  towards the low-redshift end of the data, around $z \sim 0.6$, and that the density of the LRGs drops significantly faster with redshift compared to the other samples. The eBOSS ELG distribution is at higher redshift and peaks around $z \sim 0.8$ with a significant signal density at the highest redshift $z \sim 1.0$. As the low redshift end of the intensity maps is significantly contaminated by RFI, we lose the peak of the LRG and WiggleZ sample in the cross-correlation. The total number of galaxies for the samples is significantly reduced from $N_{\rm Wig,all}=7445$, $N_{\rm LRG,all}=5632$, and $N_{\rm ELG,all}=15553$ to $N_{\rm Wig}=4815$, $N_{\rm LRG}=3281$, and $N_{\rm ELG}=8534$, respectively.

\subsection{Simulations}
In order to examine the underlying astrophysics of \hinospace-galaxy cross-correlations, we use the online service ``Theoretical Astrophysical Observatory'' (TAO\footnote{https://tao.asvo.org.au/}) to create a mock galaxy catalogue. We create the galaxy distribution using the semi-analytic galaxy formation model DARK SAGE \citep{2016MNRAS.461..859S} run on the merger trees of the Millennium simulation \citep{2006Natur.440.1137S} with box of comoving side length of $500\, {\rm Mpc}/h$. DARK SAGE is a modified version of SAGE \citep{2006MNRAS.365...11C}, which includes a pressure-based description of the atomic and molecular gas components of the cold gas based on an advanced computation of disk structure and cooling processes. DARK SAGE is calibrated to reproduce the Stellar, \hi and ${\rm H}_2$ Mass Functions as well as the fraction of \hi to stellar mass as a function of stellar mass as observed at $z=0$. For more details, we refer the reader to \citet{2016MNRAS.461..859S}. In our study, we create a lightcone with the same survey geometry covering the redshift range $0.6<z<1.0$, and the same spatial and redshift binning as the GBT data. 

We post-process the galaxy catalogue from TAO to create \hi intensity maps as well as the three optical galaxy samples. We apply the same resolution-motivated mass cut as in \citet{2016MNRAS.461..859S} and only use galaxies with $M_*>10^{8.5}M_{\rm sun}$ for our analysis. This might be a slightly conservative choice compared to, for example, \citet{2020MNRAS.493.5434S}, but the specific purpose of this simulation is to examine the \hi content of the galaxy samples rather than the universal properties of the \hi maps.  Furthermore, \citet{2020MNRAS.493.5434S}  showed that for low redshift observations, resolution effects of Millennium-based simulations are negligible for $k<1h/{\rm Mpc}$.

For the \hi intensity maps, we sum the \hi mass $M_{i,{\rm HI}}$ of all galaxies falling into the same pixel $i$ with spatial dimension $\delta \phi=\delta \theta=0.067\deg$ and the same frequency bins as the data, where we also include redshift space distortions via line-of-sight peculiar velocities of the galaxies. We transform the maps in brightness temperature using 
\begin{equation}
    T_{\rm HI}(x_i) = \frac{ 3A_{12}\hbar c^3 }{ 32\pi m_{\rm H} k_{\rm B}\nu_{\rm HI}^2} \frac{(1+z_i)^2}{H(z_i)}\frac{M_{i, \rm HI}}{V_{\rm pix}} \, ,
\end{equation}
with $\hbar$ the Planck constant, $ k_{\rm B}$ the Boltzmann constant, $m_{\rm H}$ the Hydrogen atom mass, $\nu_{\rm HI}$ the rest frequency of the \hi emission line,  $c$ the speed of light, $A_{12}$ the transition rate of the spin flip, and $V_{\rm pix}$ the co-moving volume of the pixel at mid-redshift. 
We also remove the mean temperature $\bar{T}_\hinospace$ of each map to create temperature fluctuation maps, also referred to as over-temperature maps. We then convolve the resulting maps with a Gaussian beam with ${\rm FWHM}=0.44\deg$.

Based on our galaxy lightcone catalogue, we additionally create optical, near-infrared and UV band emissions for each galaxy with the Spectral Energy Distribution (SED) module of TAO, using the Chabrier Initial Mass Function \citep{Conroy_2012}. The SED is based on the star-formation history primarily dependent on stellar mass, age, and metallicity of each galaxy. Galaxy photometry is applied after the construction of the SED. In our case, we use the SDSS filter $\{g, r, i, z\}$, and the Galex near ultra-violet filter NUV and FUV, as well as the near-infrared filter IRAC1 as an approximation for the WISE filter W1.

We apply the same observational colour cuts to the simulated lightcone to create mock galaxy samples resembling the eBOSS LRG, eBOSS ELG and WiggleZ selections, following the approach in \citet{2016MNRAS.458.3399W}. Details on the target selection are given in Appendix~\ref{app:samples}.

\begin{figure}
    \centering
    \includegraphics[width=\columnwidth]{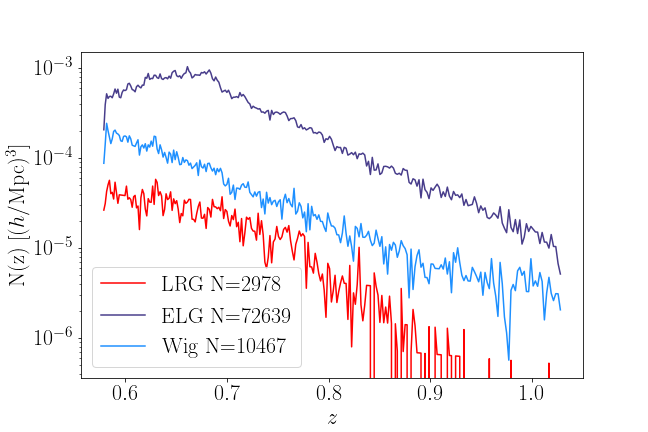}
    \caption{The galaxy density of the mock galaxy samples from the DARK SAGE simulation as a function of redshift.}
    \label{fig:Nz_sim}
\end{figure}

In \autoref{fig:Nz_sim}, we show the redshift distribution of the resulting mock galaxy samples from the semi-analytic simulation. We note that the overall galaxy numbers are off by several factors as there are many observational subtleties that can not be replicated by our approach. In addition, the eBOSS ELG-like sample peaks at slightly lower redshift around $z \sim 0.7$ compared to the actual data. However, we can see that the overall trends of the galaxy redshift distribution are present in our mock samples, and we believe that they qualitatively sample the respective galaxy types and allow us to investigate the relation between galaxy types and their \hi abundance. In this work, we use the simulation to qualitatively study the predicted \hi abundance in the galaxy samples and examine their impact on the cross-correlation power spectrum. Particularly, we investigate the non-linear shape the correlations and the amplitude of the predicted cross-shot noise. We only perform qualitative rather than quantitative comparisons between the power spectra of the semi-analytic simulation and the data.

\begin{figure}
    \centering
    \includegraphics[width=\columnwidth]{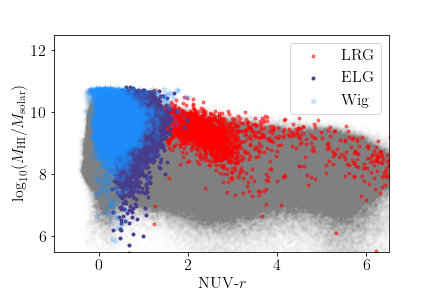}
    \caption{The \hi mass $M_{\rm HI}$ of our mock galaxy lightcone as a function of galaxy colour, $({\rm NUV}-r)$. The full light cone of $N=8.7 \cdot 10^6$ galaxies with $M_*>10^{8.5}M_{\rm sun}$ spanning $0.6<z<1.0$ is represented in grey, and the galaxy samples in coloured dots.}
    \label{fig:Colour_MHI_sim}
\end{figure}
In \autoref{fig:Colour_MHI_sim}, we present the galaxy colour to \hi mass diagram, where we use the combination of Galex-NUV and SDSS-$r$ filter to project the galaxies onto the red-blue colour scale. The NUV-$r$ colour division has been shown to be a good proxy for the star formation activity of the objects, see e.g. \cite{Cortese:2011sj}. We can see that all three samples occupy different spaces in the colour diagram with WiggleZ galaxies testing the bluest, most highly star-forming objects that are also rich in \hi gas. The ELG sample contains slightly less blue systems with lower star formation and also spanning a wider range of \hi masses. The LRG selection incorporates objects more red in colour, however, since objects are supposed to be large and luminous enough for detection at such high redshift, these are still relatively \hi rich. 

\section{Foreground Subtraction}
\label{sec:FGremoval}

\subsection{\textsc{FastICA}}

Fast Independent Component Analysis (\textsc{FastICA}) \citep{Hyvrinen1999FastAR} is one of the most popular methods for 21cm foreground cleaning and has been tested on simulated data \citep{Chapman:2012yj,Wolz:2013wna, Cunnington:2019lvb} as well as real data from the GBT \citep{Wolz:2015lwa} and LOFAR \citep{Hothi:2020dgq}. As with most foreground removal methods, \textsc{FastICA} exploits the fact that the foregrounds dominated by  synchrotron and free-free emission smoothly scale in the line-of-sight direction (frequency) \citep{2003MNRAS.346..871O,Seo_2010,PhysRevD.83.103006}, whereas the \hi signal from the Large Scale Structure follows a near-Gaussian approximation with frequency. 
We apply \textsc{FastICA} to the GBT intensity mapping data cube in order
to remove the foregrounds and non-Gaussian systematics and noise. We provide a brief summary of the method here, and refer the interested reader to \cite{Wolz:2013wna,Wolz:2015lwa} for more details. 

\textsc{FastICA} is a blind component separation method designed to divide a mixture of signals into its individual source components, commonly referred to as the ``Cocktail Party problem''. It operates on the assumption that the observed signal is composed of statistically independent sources which are mixed in a linear manner. More specifically, 
the technique solves the linear problem
\begin{equation}
\boldsymbol x= \mathbf A \boldsymbol s + \epsilon=
\sum_{i=1}^{N_{\rm{IC}}}\boldsymbol{a_i} s_i + \epsilon,
\label{eq:ica}
\end{equation}
where $\boldsymbol x$ is the mixed signal, $\boldsymbol s$ represents the $N_{\rm IC}$ independent components (ICs), and $\mathbf A$ the mixing matrix. $\epsilon$ is the residual of the analysis.  The amplitude of each IC $s_i$ is given by the mixing modes $\boldsymbol{a_i}$. \textsc{FastICA}
separates the signal into components by using the Central Limit theorem,
such that the non-Gaussianity of the probability density function of
each IC is maximized. This implies that FastICA by definition only incorporates data into $\mathbf A \boldsymbol s$ that will maximise the non-Gaussianity. The residual $\epsilon$ is obtained by subtracting the $N_{\rm IC}$ components from the original data and this should contain mostly Gaussian-like signal.

In our application of \textsc{FastICA}, the input data is of dimension $N_{\rm pix} \times N_{\nu}$ and the algorithm constructs the mixing matrix $\mathbf A$ with dimension $N_{\rm IC} \times N_{\nu}$ and the ICs $\boldsymbol s$ with dimension $N_{\rm pix} \times N_{\rm IC}$. 

\textsc{FastICA} incorporates any features with frequency correlation, such as point sources, diffuse foregrounds and non-Gaussian noise and systematics into the ICs. It also identifies frequency-localised RFI contributions with weak correlations, as they usually exhibit strong non-Gaussian spatial features. The residual of the component separation should, in theory, only contain the \hi signal and the Gaussian telescope noise.

The number of ICs ($N_{\rm IC}$) used in the component separation is a
free parameter and can not be determined by \textsc{FastICA}. 
In the following sub-sections, we carefully examine the sensitivity of the 
foreground-subtracted data to different choices of $N_{\rm IC}$, ensuring that our results do not depend on this choice.

\subsection{Transfer Function}
\label{subsec:transfer}
Foreground subtraction with \textsc{FastICA} and its applications to simulations has been thoroughly investigated by many studies \citep{Wolz:2013wna, Alonso:2014dhk, 2020MNRAS.495.1788A, Cunnington:2020njn}, but the vast majority of simulations published to date have been highly idealised and do not included any instrumental effects other than Gaussian noise. In this idealised setting, \textsc{FastICA} has been found to very effectively remove foregrounds for low numbers of ICs starting from $N_{\rm IC}=4$. We note that these numbers also depend on the sophistication of the foreground models, for example, see \citet{Cunnington:2020njn} for  $N_{\rm IC}>4$ in the case where polarisation leakage is included in the simulations. 

\textsc{FastICA} applied to systematics dominated data can effectively remove non-Gaussian and anisotropic systematics \citep{Wolz:2015lwa}, as well as the astrophysical foregrounds. This means that for increasing $N_{\rm IC}$, the algorithm incorporates more subtle signals as well as more local features into the components. This can significantly reduce the presence of noise and systematics in the data, however, it could also lead to \hi signal loss.

In the following, we investigate the signal loss for different numbers of $N_{\rm IC}$ in the presence of systematics and use the methodology presented in \cite{Switzer_2015} to construct the transfer function to correct for \hi signal loss.
In absence of a telescope simulator for the (unknown) systematics, we obtain the transfer function by injecting mock \hi signal from simulations into the observed maps before foreground removal. We then process the combined maps with \textsc{FastICA}, and determine the \hi signal loss by cross-correlating the cleaned maps with the injected \hi simulation. In order to reduce noise, we use the average of 100 \hi realisations and we also subtract the cleaned GBT data from the combined data before cross-correlating with the injected signal.

\noindent We describe the process in detail below:
  
 \begin{itemize}
    \item We create $N_m=100$ mock simulations $m_i$ of lognormal halo distributions using the python package \textsc{powerbox} \citep{Murray2018} with a halo mass limit of $M_{h,{\rm min}}=10^{12.3}M_\odot/h$. 
    \item We populate each dark matter halo with a \hi mass following a simple \hi halo mass relation as in \cite{2019MNRAS.484.1007W}.
    \item We grid the \hi mass of each halo to the same spatial and frequency resolution as the GBT data at median redshift $z \approx 0.8$.
    \item We convert the \hi grid into brightness temperature $T_{\rm HI}$ using \autoref{eq:thi}, re-scale the overall averaged temperature to the same order of magnitude as the theory prediction with $\Omega_{\rm HI}=0.5\times 10^{-3}$, and convolve the data with a constant, symmetric Gaussian beam with ${\rm FWHM}=0.44\deg$. 
    \item We add each mock \hi brightness temperature realisation $m_i$ to each GBT season 
    $j \in \{A, B, C, D\}$ of the GBT data to create combined cubes $(d_j+m_i)$.
    \item We run \textsc{FastICA} with $q$ number of independent components on each sub-dataset as ${\rm ICA}_q(d_j+m_i)$, where $q \in \{ 4, 8, 20, 36\}$.
    \item We subtract the original, cleaned GBT data cube to obtain the cleaned mock simulations $\tilde m_{qi}^j ={\rm ICA}_q(d_j+m_i) - {\rm ICA}_q(d_j)$ for each realisation $i$, each GBT season $j$ and each choice of foreground removal $N_{\rm IC} = q$.
 \end{itemize}
 
A comparison of the amplitudes and shapes of the auto-power spectrum of the foreground cleaned injected mock $\tilde m_{qi}^j$ and auto-power spectrum of the original mock $m_i$ measures the \hi signal loss of the power spectrum through the foreground removal. However, in this study, we are interested in quantifying the \hi signal loss through foreground subtraction on the cross-correlation power spectrum with galaxy surveys. In order to approximate this effect, we examine the cross-power spectrum of the foreground removed mock $\tilde m_{qi}^j$ with the original mock $m_i$, where the original mock acts as approximate of the galaxy field with cross-correlation coefficient equal to unity. We define the signal loss function $\Delta$ per season $j$ for different $q=N_{\rm IC}$ averaged over all realisations as 
 
 \begin{equation}
         \Delta^{j}_{q}(k)=\frac{\sum_i^{{N_m}}P(\tilde m_{q,i}^j, m_i)(k) }{\sum_i^{N_m}P(m_i)(k)} \, .
\end{equation}
In an ideal situation without any signal loss, $\Delta^{j}_{q}(k)$ is equal to unity across all scales. Note that here, $\Delta$ is defined as the \hi signal loss function on the \hinospace-galaxy cross-correlation.

 \begin{figure}
    \centering
    \includegraphics[width=\columnwidth]{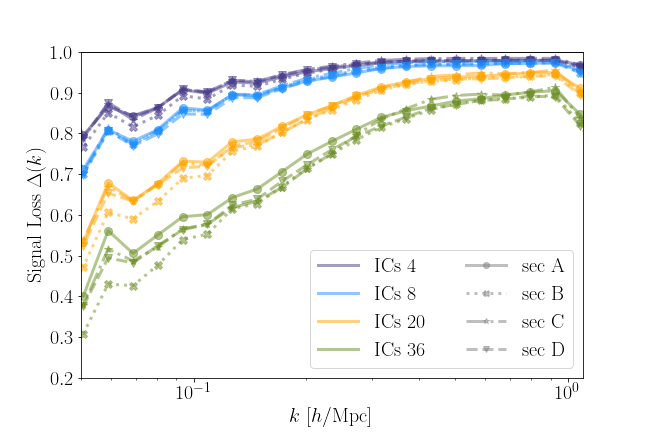}
    \caption{The signal loss function $\Delta(k)$ for the foreground subtraction with \textsc{FastICA} for different numbers of ICs $N_{\rm IC}$. Note that $\Delta=0.8$ means $20\%$ signal loss. We show the the individual seasons $\rm \{A, B, C, D\}$ to highlight the sensitivity of the transfer function to the individual season-dependent systematics.}
    \label{fig:signalloss}
\end{figure}

In our analysis, the signal loss is corrected via the transfer function of the cross-correlation  defined as $\Theta^j_q=(\Delta^j_q)^{-1}$. We show the signal loss function in \autoref{fig:signalloss}. For all tested $N_{\rm IC}$, there is some significant degree of signal loss ranging between $10\%-50\%$ on the largest scales $k<0.1 \, h{\rm Mpc^{-1}}$. This can be explained considering the survey geometry, as these scales are mostly tested by line-of-sight modes which are highly affected by diffuse foreground subtraction. Even for increasing numbers of ICs in the subtraction, the transfer function converges towards unity on smaller scales. However, for very high number $N_{\rm IC}=36$, there is signal loss on all scales of the power spectrum. Note that the divergent behaviour from $k>1h{\rm Mpc}^{-1}$ is due to the effect of the beam on these scales, and they are not considered in our final analysis. We can see that in general, the amplitude of the transfer function of season B is somewhat higher than the others, which suggests that this season might suffer more from systematic effects.

\section{Power spectrum Results}
\label{sec:PSresults}

We use the inverse-noise weighted power spectrum estimator as described in \cite{Wolz:2015lwa}. For the cross-correlation of two tracers $a$ and $b$, that is:
\begin{equation}
    \hat P^{ab}(\vec k_l)=\frac{V \mathrm{Re}\{ \tilde \delta^a(\vec k_l)\cdot
\tilde\delta^b(\vec k_l)^*\} 
}{\sum_{j=1}^{N_{\rm pix}} w^a(\vec x_j)\cdot  w^b(\vec x_j)} \, ,
\label{eq:PScross}
\end{equation}
with $\tilde \delta$ the Fourier transform of the weighted density field $w(\vec x_j)\delta(\vec x_j)$ of the tracer, $N_{\rm pix}$ the total number of pixels, $ w(\vec x_j)$ the weighting function, and $V$ the survey volume. For \hi intensity maps, $w(\vec x_j)$ is given by the inverse noise map of each season. For galaxy surveys, the total weighting factor is $w(\vec x_j)=W(\vec x_j)w_{\rm opt}(\vec x_j)$, where $w_{\rm opt}(\vec x_j)$ is given by optimal weighting function $w_{\rm opt}(\vec x_i)=1/(1+W(\vec x_i)\times \bar N P_0)$, with $P_0=10^3 h^{-3}\rm{Mpc}^3$, and the selection function $W(\vec x_j)$. We derive the selection function for each sample from binning the random catalogues. The redshift evolution of these is shown as dashed lines in \autoref{fig:Nz_gal}, and the spatial footprint in \autoref{fig:Sel_gal}. We note, that we do not use any additional weighting functions for the galaxy power spectrum. 

\autoref{eq:PScross} holds for \hinospace-auto, galaxy-auto, as well as \hinospace-galaxy correlations. For galaxy power spectra, we additionally remove the shot noise weighted by the selection function as described in \citet{blake2011}. The 1-d power spectra $\hat P(k)$ are determined by averaging all modes with $k = |\vec k|$ within the $k$ bin width.

In the following, we use $\hat P$ to indicate the estimated power spectrum, and $P$ for the theory prediction. All power spectra are estimated using the redshift range $0.62<z<0.95$ with $N_{\nu}=190$, and spatial resolution $N_{\rm RA}=187$ and $N_{\rm DEC}=89$. We use the flat sky approximation at mid-redshift $z=0.78$, resulting in a volume of $V=4.2\cdot 10^7 ({\rm Mpc}/h)^3$. Note, that we do not correct for gridding effects with our power spectrum estimator since the power spectrum is dominated by the beam from $k \sim 1 \, h{\rm Mpc}^{-1}$.

\subsection{\hi Power Spectrum}
\begin{figure*}
    \centering
    \includegraphics[width=0.8\textwidth]{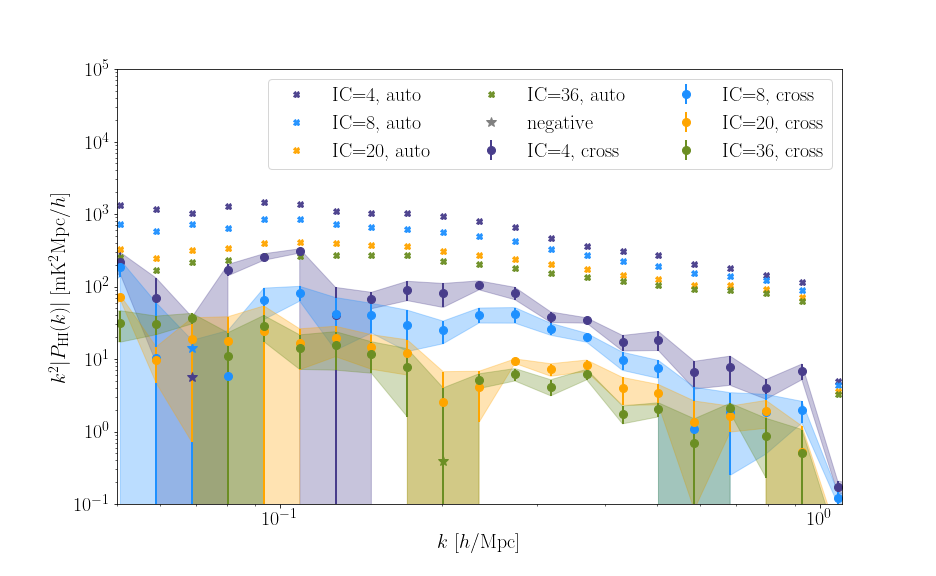}
      \caption{The absolute value of the \hi power spectrum of the GBT intensity maps for different number of ICs in the foreground subtraction. All power spectra are transfer function corrected. We show the auto-correlation between the seasons marked with crosses, and the season cross-correlation with circles. There are a few negative data points (indicated by stars), which demonstrate the high noise on the measurements. Note that these measurements are about an order of magnitude higher than theory predictions and should be treated as upper limits which is in agreement with \citep{Switzer_2015}.}
    \label{fig:autoPS_HI}
\end{figure*}

In this section, we present the \hi power spectrum to visualise the impact of the foreground subtraction and the transfer function.
In \autoref{fig:autoPS_HI}, we show the \hi power spectrum in auto-correlation $\hat P_{\rm HI}^{i}$ for each season $i$, as well as the cross-correlation between the seasons $\hat P_{\rm HI}^{ij}$ for all investigated numbers of ICs $N_{\rm IC}\in \{4, 8, 20,36\}$. We present the \hi power spectrum with foreground subtraction correction, where we use $\Theta^2_i$ as an approximation to correct the auto-power spectrum $i$, and $\Theta_i\Theta_j$ to correct for the cross-season correlation $ij$. 
As expected, the auto-power spectrum is dominated by instrument noise whose amplitude is higher than the \hi signal. Unlike other subtraction techniques like PCA, \textsc{FastICA} cannot remove and mitigate effects of Gaussian telescope noise. Hence, $P_{\rm HI}^{i}$ can be used as an estimate for the noise present in the data and we use the averaged auto-power spectrum $\hat P_{\rm HI, q}^{\rm auto}(k) = \sum_i^4 \hat P_{\rm HI, q}^{i}(k) /4$ to estimate the error bars on the \hi power spectrum as:
\begin{equation}
   \sigma_{\rm HI,q}(k) = \hat P_{\rm HI, q}^{\rm auto}(k)/\sqrt{2N_{\rm modes}} \, ,
   \label{eq:shi}
\end{equation}
with $N_{\rm modes}$ the number of $k$ modes sampled in the survey volume, and $q$ the number of ICs, $N_{\rm IC}$. As we use the auto-correlation between seasons as proxy for the noise on the \hi power spectrum, an extra scaling of $\sqrt{2}$ is applied to the error between seasons.

Another way to estimate the noise directly from the data, is using the scatter between cross-season power spectra as noise estimate, and we find that the standard deviation of cross-season is the same order of magnitude as the auto-power spectrum, however, given the limited number of independent seasons, the auto-power spectrum is much less sensitive to sampling variance. For a comparison of these two approaches on the GBT data, please refer to Fig 8 of \citep{Wolz:2015lwa}.

The cross-season power spectra contain a few negative data points, which are indicated by stars in \autoref{fig:autoPS_HI}. This is the result of the high noise properties in the map which can dominate certain scales. 

For the cross-season power spectra, we can see that the amplitude of the spectra is starting to converge for increasing number of $N_{\rm IC}$ on all scales. We are therefore confident that these two choices of ICs in the foreground subtraction are removing sufficient foregrounds. We use $N_{\rm IC}=20$ as a conservative choice with minimal \hi signal loss, and possibly higher residual systematics and noise. Whereas $N_{\rm IC}=36$ is a more assertive choice in the subtraction resulting in lower noise properties with higher levels of \hi signal loss.

\subsection{Galaxy Power spectrum}
\begin{figure}
    \centering
    \includegraphics[width=\columnwidth]{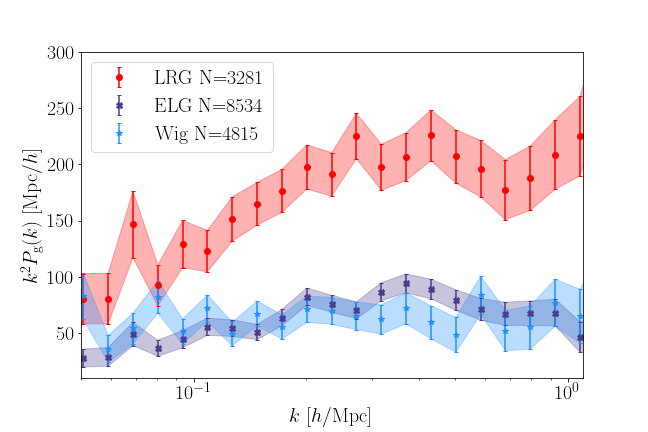}
    \includegraphics[width=\columnwidth]{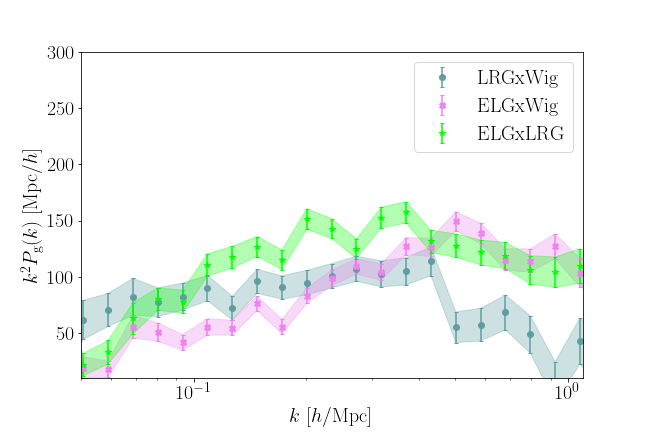}

    \caption{The galaxy power spectrum of the eBOSS LRG, eBOSS ELG and WiggleZ samples. \emph{Top}: The auto-power spectra of the individual samples, with higher amplitude in the LRG sample and similar amplitudes of the ELG and WiggleZ samples, reflecting the different biases of the samples. \emph{Bottom:} The cross-correlation between galaxy samples. We observe a drop in small scale amplitude for the LRG-WiggleZ correlation. }
    \label{fig:autoPS_gal}
\end{figure}

In \autoref{fig:autoPS_gal}, we show the galaxy power spectra $\hat P_{\rm g}(k)$ of our samples in auto- as well as cross-correlation. Note that our power spectrum estimator is not optimised for galaxy surveys and we do not use the galaxy power spectra for a quantitative analysis. Only the auto-galaxy power spectra are shot noise removed, as we do not assume a sample overlap between galaxy surveys.

The error bars on the auto-correlation are estimated as 
\begin{equation}
    \sigma_{\rm g}(k) = \frac{1}{\sqrt{N_{\rm modes}}}\left(\hat P_{\rm g}(k)+\frac{1}{n_{\rm g}}\right) \, ,
    \label{eq:sgal}
\end{equation}
where $N_{\rm modes}$ is again the number of independent $k$ modes in the survey volume, and $n_{\rm g}$ is the galaxy density of the samples, computed as $n_{\rm g} = N_{\rm g}/V$, with $N_{\rm g}$ the number of galaxies and $V$ the survey volume. The cross-galaxy error bars are estimated as
\begin{equation}
    \sigma_{\rm g}^{ij}(k) = \frac{1}{\sqrt{2N_{\rm modes}}}\sqrt{\hat P_{\rm g}^{ij}(k)^2+\left(\hat P_{\rm g}^i(k)+\frac{1}{n_{\rm g}^i}\right) \left(\hat P_{\rm g}^j(k)+\frac{1}{n_{\rm g}^j}\right) } \, .
\end{equation}
In the upper panel of \autoref{fig:autoPS_gal}, we can see that the ELG and WiggleZ samples are similarly biased across scales, with tentatively an opposite trend in the scale-dependent behaviour. This result is in agreement with theory, as the WiggleZ and ELG samples trace similar populations of galaxies. The bias of the LRG sample is significantly higher, which is again as expected as this sample traces more quiescent, early-type objects in denser environments. 

The lower panel of \autoref{fig:autoPS_gal} shows the cross-correlation between the galaxy samples, similarly to \cite{Anderson_2018}. The idea being that the bluer, star-forming samples (ELG and WiggleZ) trace the dark matter in a similar manner to \hinospace, therefore the shape of the blue-red correlation power spectrum could also be used as a qualitative estimate of the \hinospace-LRG cross power spectrum. In our data, most notably, the WiggleZ-LRG power spectrum exhibits a drop in amplitude for smaller scales which is not seen for the other two spectra.

\subsection{\hinospace-Galaxy Power Spectrum}
\begin{figure}
    \centering
    \includegraphics[width=\columnwidth]{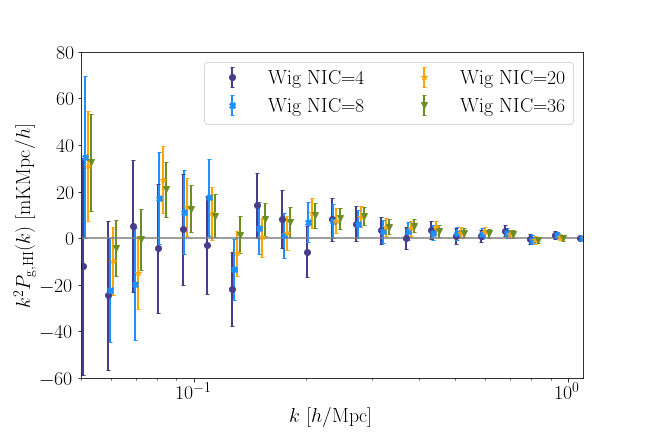}
    \includegraphics[width=\columnwidth]{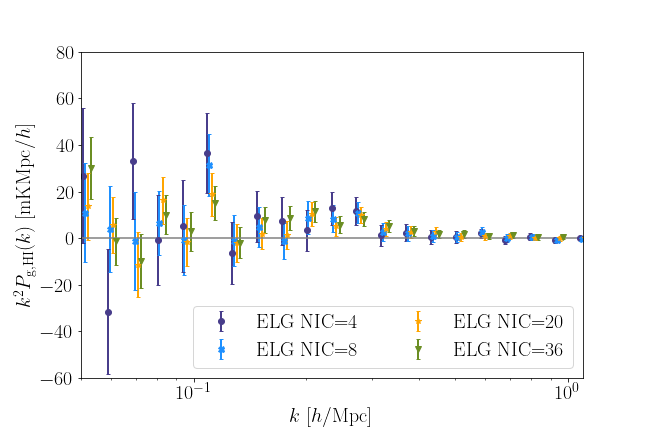}
    \includegraphics[width=\columnwidth]{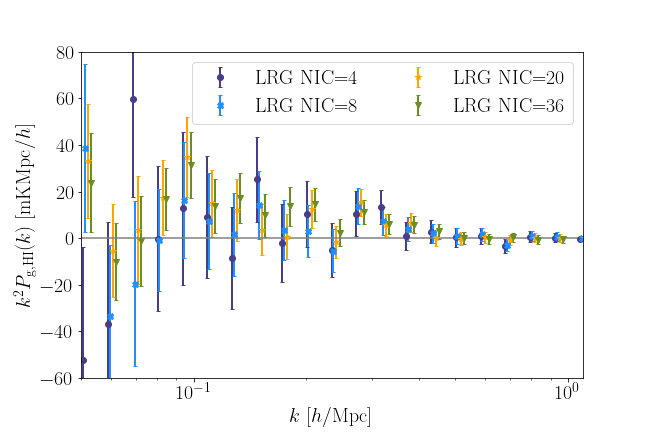}

    \caption{The GBT \hi intensity mapping cross-correlation with the galaxy samples for different numbers of ICs in the foreground subtraction. Note that all power spectra were estimated at the same $k$, and the staggered $k$ values in the plots are for illustration purposes only.  \emph{From top to bottom:} \hinospace-WiggleZ, \hinospace-ELG, and \hinospace-LRG cross-correlation power spectrum. }
    \label{fig:crossPS_HIgal_IC}
\end{figure}

In \autoref{fig:crossPS_HIgal_IC}, we present the \hinospace-galaxy cross-power spectra in absolute power for the three galaxy samples and different numbers of ICs in the foreground subtraction. The error bars on these power spectra are determined by the errors
on the galaxy sample, see \autoref{eq:sgal}, and the \hi data, see \autoref{eq:shi}, combined as
\begin{equation}
    \sigma_{\rm g, HI}^{q}(k) = \frac{1}{\sqrt{2N_{\rm modes}}}\sqrt{\hat P_{\rm g, HI}^{q}(k)^2+ \hat P_{\rm HI}^q(k)\left(\hat P_{\rm g}(k)+\frac{1}{n_{\rm g}}\right) } \, ,
    \label{eq:sgHI}
\end{equation}
with $q$ the number of ICs $\{ 4, 8, 20, 36\}$. We note that the \hi data errors dominate the total cross-power error budget. We discuss errors and covariances in more detail in \autoref{analysistests} and \autoref{appb}.
\begin{figure}
    \centering
    \includegraphics[width=\columnwidth]{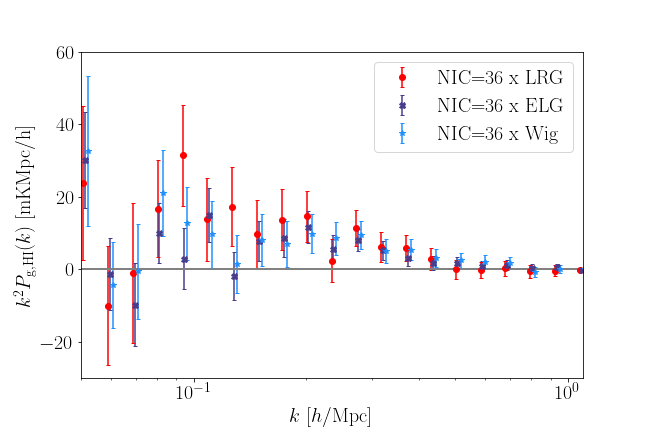}
   
    \caption{The GBT \hi intensity mapping cross-correlation with the galaxy samples in comparison. Note that all power spectra were estimated at the same $k$, and the staggered $k$ values in the plots are for illustration purposes only.}
    \label{fig:crossPS_HIgal}
\end{figure}

We can see in all three panels of \autoref{fig:crossPS_HIgal_IC}, that the amplitude of the cross-power signal is not very sensitive to the foreground removal parameters within the error bars. We do not observe a drop in amplitude with increasing numbers of ICs, and we are confident that we correctly account for \hi signal loss with our transfer function, particularly, within the large errors of the GBT data. Generally, as the amplitude of the noise of the GBT data is decreased with increasing $N_{\rm IC}$, the detection of the signal becomes more statistically significant and the error bars decrease with increasing components removed. In \autoref{fig:crossPS_HIgal}, we show the cross-correlation of the three galaxy samples for fixed $N_{\rm IC}=36$ in comparison.

The GBT-WiggleZ cross-correlation in the upper panel of \autoref{fig:crossPS_HIgal}, is detected for both $N_{\rm IC}=20,36$ on scales $0.1 < k < 0.8 \, h{\rm Mpc}^{-1}$. Qualitatively, the middle panel showing the amplitude of the GBT-ELG correlation looks very similar, but the detection seems more noise dominated on the larger scales, around $k\approx 0.1h{\rm Mpc}^{-1}$. 
The GBT-LRG correlation shown in the lowest panel demonstrates a detection of the signal for $N_{\rm IC}=36$. At the smallest scales around $k \sim 1 \, h{\rm Mpc}^{-1}$, the amplitude of the correlation signal drops off and the power spectrum is highly noise dominated. \cite{Anderson_2018} reported a drop in amplitude in the cross-correlation of the Parkes \hi intensity maps with the red sub-sample 2dF galaxies. However, the signal-to-noise ratio of the GBT-LRG measurements is not large enough to confirm this trend.

The cross-correlation of WiggleZ-LRG galaxies as shown in \autoref{fig:autoPS_gal} supports that this would be an expected result for our data. The negligible power of the correlation of the \hi intensity maps with the LRG galaxy sample on small scales, implies that the LRG galaxies that contribute to these scales are \hi deficient. The power spectrum signal on these scales originates from galaxy pairs most likely part of the same halo in a dense cluster environment. The \hi deficiency of these types of quiescent galaxies has been predicted in theory and observed for the local Universe\citep{Reynolds:2020tm}. Our work is an indicator for this trend for cosmological times.

We will make more quantitative estimates for the significance of the detections when we present our derived \hi constraints in \secref{sec:HIconstraints}.

\subsection{Comparison to Simulations}
We use our simulations for qualitative interpretation of our results. We use the same redshift range with $\bar z\approx 0.78$ to estimate the power spectra of our mock data, however, we do not mask the edges of the data which results in a bigger volume of $V=4.8\cdot 10^7 ({\rm Mpc}/h)^3 $. We do not include any noise and instrumental effects in this simulation suite as we focus on understanding the implication from galaxy evolution on the cross-correlation signal.

\begin{figure}
    \centering
     \includegraphics[width=\columnwidth]{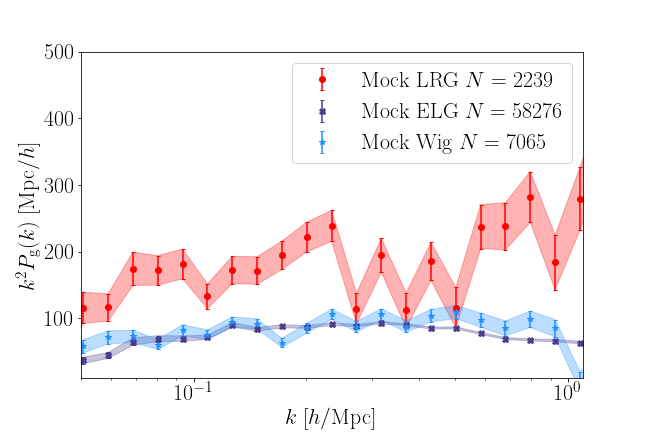}
    \includegraphics[width=\columnwidth]{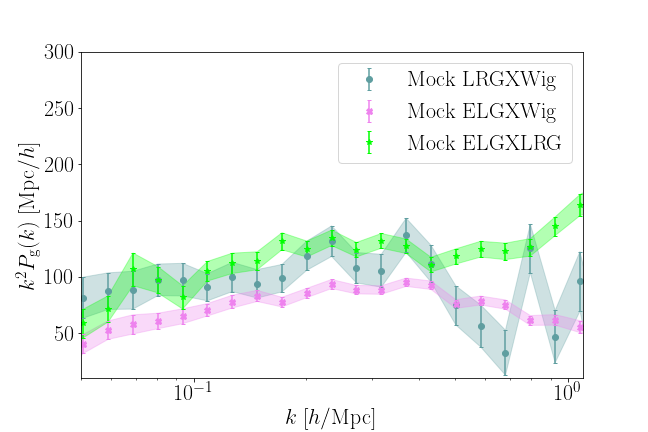}
    \includegraphics[width=\columnwidth]{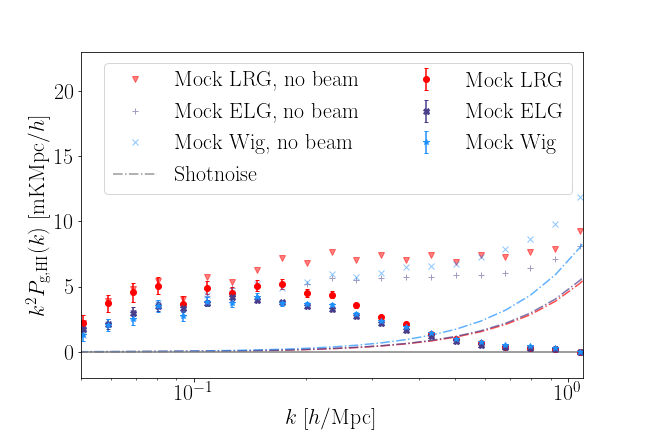}
    \caption{The power spectra of our simulation suite. \emph{Top:} The auto-galaxy power spectra of the three galaxy samples. The mock-ELG and WiggleZ power spectra are of similar amplitude, whereas the mock-LRG exhibits a higher bias, consistent with the data. \emph{Middle:} The cross-galaxy power spectra of the mock samples. Similarly to the data, we see a possible drop in amplitude on smaller scales for the LRG-WiggleZ correlation. \emph{Bottom:} The \hinospace-galaxy cross-correlation, beam-convolved and with no beam to demonstrate the effect of the cross-shot noise. The dashed-dotted lines indicate the shot noise amplitude. }
    \label{fig:mockPS}
\end{figure}

In \autoref{fig:mockPS} from top to bottom, we show the power spectra for the galaxy samples, the cross-galaxy and the \hinospace-galaxy correlations. The shapes and amplitudes of the galaxy power spectrum are comparable to the data power spectrum. We presume that the fluctuations of the mock LRG sample are due to the low galaxy density. The cross-galaxy power spectra are comparable to the data measurements, with a drop in amplitude at smaller scales $k>0.8 \, h{\rm Mpc}^{-1}$.

In the bottom panel of \autoref{fig:mockPS} we show the resulting mock \hinospace-galaxy cross-correlation. We note that the overall amplitude is lower than the data due to a lower $\Omega_{\rm HI}$ than data measurements suggest. The simulations predict the amplitude of all power spectra at the same level of magnitude. We show the beam-convolved mock as well as a unconvolved power spectrum, to demonstrate the effect of the \hi shot noise, as predicted in \cite{2017MNRAS.470.3220W}. The amplitude of the cross-shot noise is proportional to the ensemble averaged \hi mass of the respective galaxy sample. Our simulation predicts the highest shot noise amplitude for the \hinospace-WiggleZ correlation, and very similar levels for both eBOSS samples. However, on the scales unaffected by the GBT telescope beam, the shot noise does not have a measurable effect, in particular when considering the signal-to-noise ratio of our data. Notably, we do not find a drop in amplitude of the \hinospace-LRG correlation. This could suggest, that the drop could be caused by an unknown observational effect, which we were unable to identify with our tests given the large uncertainties of the data, or, alternatively, that our selection of mock LRG galaxies or the model itself misses some features and our mock sample can not fully represent the data. We hope to investigate this interesting feature in future work with less noise-dominated \hi intensity maps.

\subsection{Analysis tests}
\label{analysistests}
We perform several tests of our analysis pipeline listed in this section. For these tests, we examine the covariance matrix of the mock data computed as 
\begin{equation}
     \mathbf C_q= C_{q}(k_i, k_j)=\sum_m^{N_m} \frac{(P_m^q(k_i)-\bar{P}^q(k_i)) (P_m^q(k_j)-\bar{P}^q(k_j)}{N_m}
     \label{eqcovariance}
 \end{equation}
 where the number of independent components $q = \{4,8,20,36\}$, $\bar P$ the averaged power spectrum over all realisations, and $N_m$ the number of realisations. We can derive an estimate for error bars from the diagonal as 
 $\sigma_{i}^q=\sqrt{\mathbf C^q_{ii}}$. Figures of the resulting covariance matrices and tests can be found in \autoref{appb}.

\begin{itemize}
    \item Mode correlation from \textsc{fastICA}: We derive the covariance of the data to determine the statistical independence between $k$ bins. We use the power spectra $P(\tilde m_{q,i}^j, m_i)(k)$ of the foreground-subtracted lognormal simulations $\tilde m_{q,i}^j$ with the original simulation $m_i$, and compute the covariance matrix. We find no significant off-diagonal correlations between the modes $0.05< k <0.8 \, h{\rm Mpc}^{-1}$ considered in our analysis. We also compute the errors from the diagonal of the inverted covariance matrix to determine the additional error introduced from the foreground removal. We find that this contribution is more than 2 orders of magnitude lower than the analytical errors based on noise and cosmic variance as determined by \autoref{eq:shi}. We therefore can safely neglect this contribution in the present analysis.
   \item Randoms null test: We correlate the GBT sub-season data with the $N_m=100$ random WiggleZ catalogues used to derive the selection function. As expected, we find a signal consistent with zero within the error bars. We also derive the covariance matrix from the mocks and find that the error bars $\sigma_{\rm cov}$ are in agreement with the empirically derived $\sigma_{\rm g,HI}$ in \autoref{eq:sgHI}.
   
     \item Shuffled null test: We correlate the GBT sub-season data with the three galaxy samples which are each re-shuffled in redshift to remove the correlation. As expected, we find all signals consistent with zero within the error bars.
\end{itemize}

\section{\hi constraints}
\label{sec:HIconstraints}
Here, we are present our derived \hi constraints from the cross-correlation power spectra analysis (summarised in \autoref{tab:constraints}). Before doing so, we briefly review the findings of \citet{Masui:2012zc}, who measured the GBT maps cross-correlation with the WiggleZ 15hr and 1hr fields. Fitting in the range of scales
$0.05 \, h{\rm Mpc}^{-1} < k < 0.8 \, h{\rm Mpc}^{-1}$, they found $10^3\Omega_\hinospace b_\hinospace r = 0.40 \pm 0.05$ for the combined,  $10^3\Omega_\hinospace b_\hinospace r = 0.46 \pm 0.08$ for the 15hr field and $10^3\Omega_\hinospace b_\hinospace r = 0.34 \pm 0.07$ for the 1hr field (which is the one we are considering in this paper). 
For a more restrictive range of scales, their combined measurement was $10^3\Omega_\hinospace b_\hinospace r = 0.44 \pm 0.07$. Note that \citet{Masui:2012zc} used Singular Value Decomposition (SVD) for their foreground removal, but we use \textsc{FastICA} here following \citet{Wolz:2015lwa}. Our transfer function construction methods are identical. 
We note that the errors quoted are statistical, and \citet{Masui:2012zc} also estimated a $\pm 0.04$ systematic error representing their $9\%$ absolute calibration uncertainty. We will adopt the same systematic error in our analysis.

In this paper we will explore different ranges of scales, by performing fits for three cases: \textbf{Case I}, with $0.05 \, h{\rm Mpc}^{-1} < k < 0.8 \, h{\rm Mpc}^{-1}$. \textbf{Case II}, with $0.05 \, h{\rm Mpc}^{-1} < k < 0.45 \, h{\rm Mpc}^{-1}$, and \textbf{Case III}, with $0.05 \, h{\rm Mpc}^{-1} < k < 0.35 \, h{\rm Mpc}^{-1}$. 
Considering different ranges of scales is motivated by the fact that, while small scales (high $k$) contain most of the statistical power of the measurement, the beam and model of non-linearities become less robust as $k$ increases.

\begin{figure}
    \centering
    \includegraphics[width=\columnwidth]{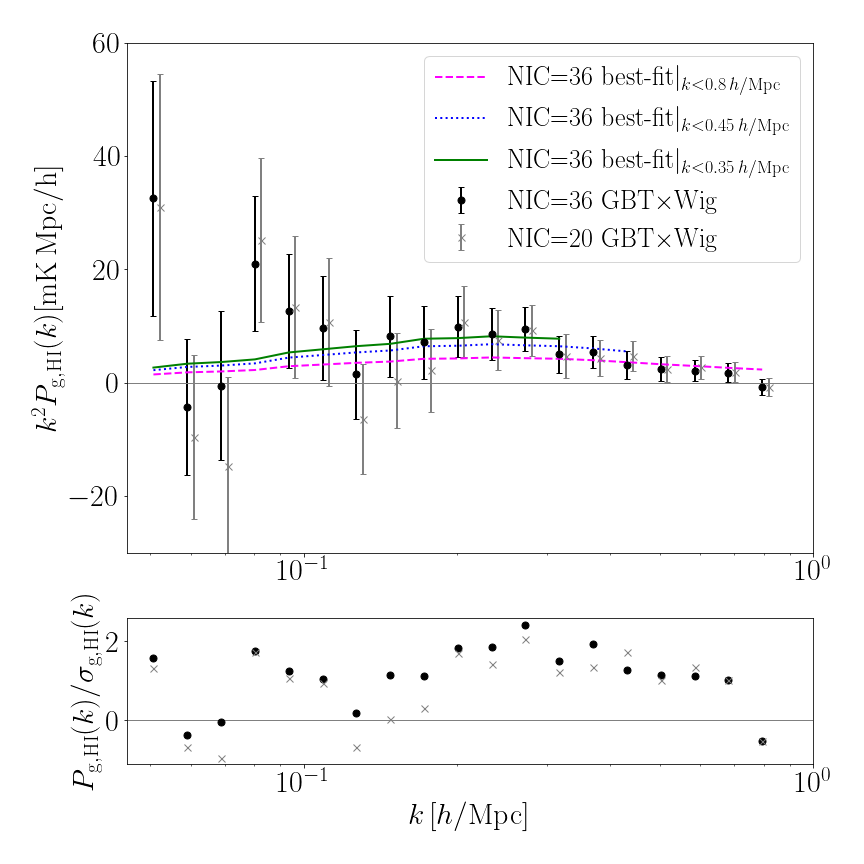}
    \caption{\emph{Top}: The measured GBT-WiggleZ cross-correlation power spectrum. We show two cases with $20$ and $36$ Independent Components used in \textsc{FastICA} for the \hi maps foreground cleaning, corrected with the corresponding transfer functions. We also show the best-fit models from \autoref{tab:constraints} (Cases I, II, and III) for $N_{\rm IC}=36$. \emph{Bottom}: A null diagnostic test plotting the ratio of data and error.}
    \label{fig:GBTxWig}
\end{figure}

\begin{figure}
    \centering
    \includegraphics[width=\columnwidth]{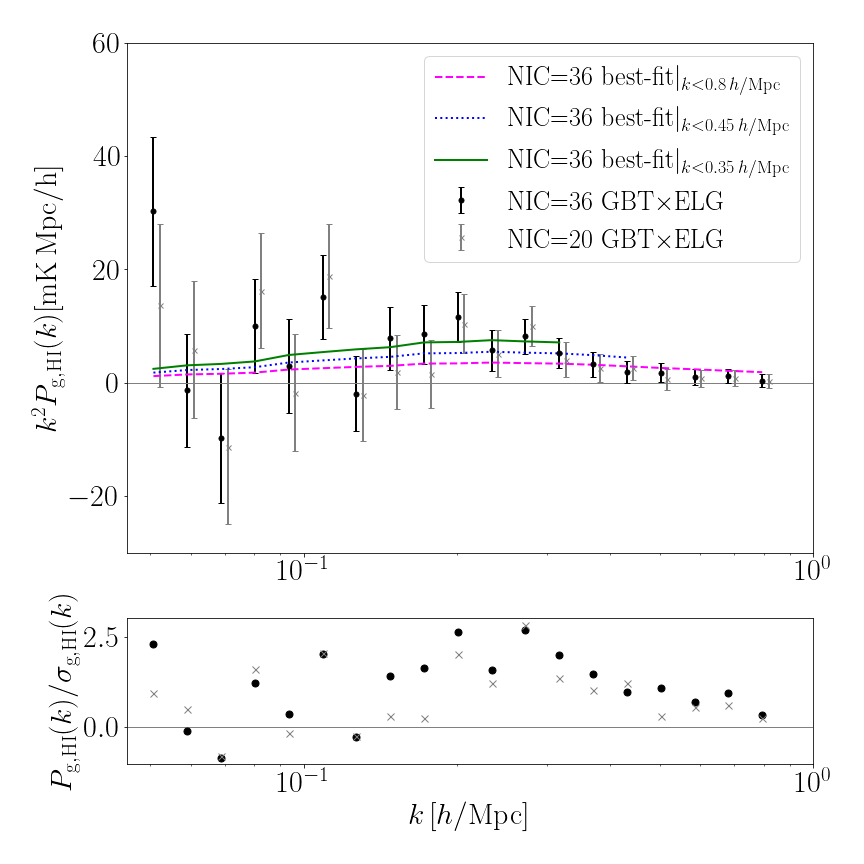}
    \caption{\emph{Top}: The measured GBT-ELG cross-correlation power spectrum for $N_{\rm IC}=20, 36$. We also show the best-fit models from \autoref{tab:constraints} (Cases I, II, and III) for $N_{\rm IC}=36$. \emph{Bottom}: A null diagnostic test plotting the ratio of data and error.}
    \label{fig:GBTxELG}
\end{figure}

\begin{figure}
    \centering
    \includegraphics[width=\columnwidth]{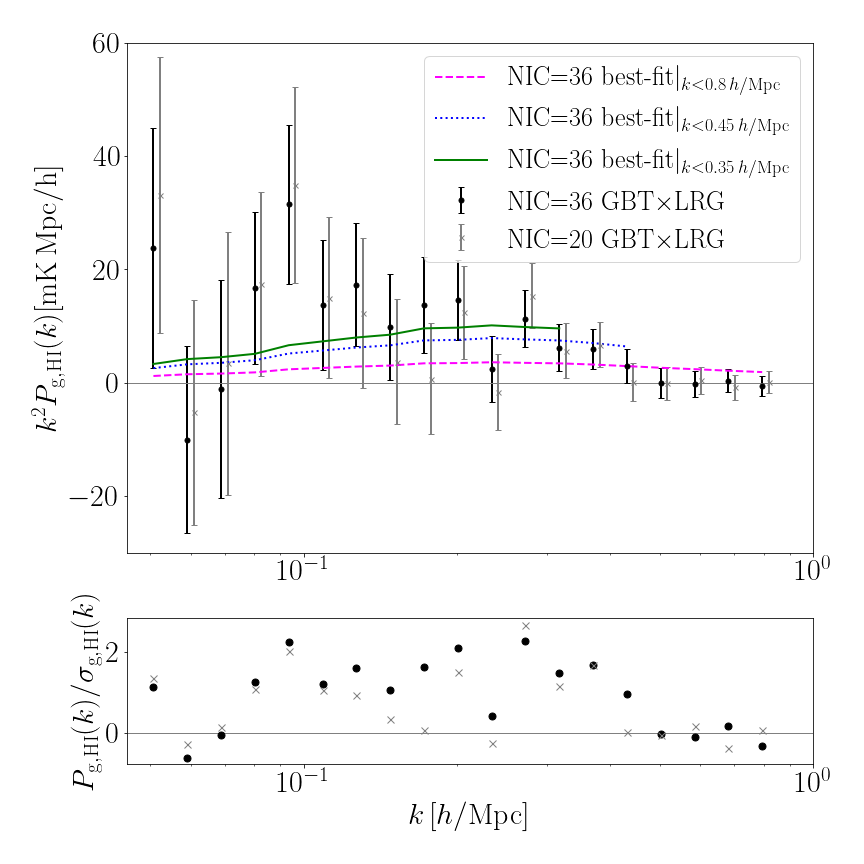}
    \caption{\emph{Top}: The measured GBT-LRG cross-correlation power spectrum for $N_{\rm IC}=20, 36$. We also show the best-fit models from \autoref{tab:constraints} (Cases I, II, and III) for $N_{\rm IC}=36$. \emph{Bottom}: A null diagnostic test plotting the ratio of data and error.}
    \label{fig:GBTxLRG}
\end{figure}

In \autoref{fig:GBTxWig} we show the measured GBT-WiggleZ power spectrum, concentrating on the results with $N_{\rm IC}=20,36$. 
In the bottom panel, we perform a simple \emph{null diagnostic test} by plotting the ratio of data and error. This shows that most of the measurements in the range of scales with high signal-to-noise ratio are more than $1\sigma$ positively away from $0$. 
For our fiducial IC=36 results for Case I, corresponding to the same range of scales considered in \citet{Masui:2012zc}, the detection significance is estimated to be $4.4\sigma$ (we note that in \citet{Masui:2012zc} this was found to be $7.4\sigma$ but for the combined 1hr and 15hr fields observations). 
We show similar plots for the GBT-ELG and GBT-LRG cross-correlations in \autoref{fig:GBTxELG} and \autoref{fig:GBTxLRG}, respectively. We note that our null tests suggest that the GBT-LRG detection is the most tentative of the three. Indeed, estimating the detection significance for GBT-ELG and GBT-LRG, we find $4.5 \sigma$ and $2.9 \sigma$, respectively, for Case I. In \autoref{tab:constraints} we show the detection significance for $N_{\rm IC}=36$ for all Cases.
We see that the detection significance for the GBT-LRG cross-correlation considerably improves when considering the restricted ranges of scales, Cases II and III. 

To relate the measured power spectra with a theory model and derive the \hi constraints, we use \autoref{eq:thi} to express the mean 21cm emission brightness temperature $T_{\hinospace}$ as a function of $\Omega_{\hinospace}$.
We observe the brightness contrast, $\delta T = T_{\hinospace}\delta_\hinospace$. We also assume that the neutral hydrogen and the optical galaxies are biased tracers of dark matter, but we also include a galaxy-\hi stochastic correlation coefficient $r_{\hinospace,{\rm opt}}$. To compare the theoretical prediction with the measurements, we follow a procedure similar to the one described in \citet{Masui:2012zc}:
\begin{itemize}
    \item We assume a fixed Planck cosmology \citep{Ade:2015xua}.
    \item We assume a known galaxy bias $b_{\rm opt}$ at the mean redshift $z\simeq 0.8$, with ${\rm opt}$ corresponding to WiggleZ \citep{blake2011}, eBOSS ELGs, and eBOSS LRGs \citep{Alam:2020sor} depending on the galaxy sample we cross-correlate the \hi maps with. That is, $b_{\rm Wig} = 1.22$, $b_{\rm ELG}=1.4$, $b_{\rm LRG}=2.3$.
    \item We include non-linear effects to the matter power spectrum $P_{\rm m}(k)$ using \texttt{CAMB} \citep{Lewis:1999bs} with \texttt{HALOFIT} \citep{Smith:2002dz,Takahashi:2012em} and also include (linear) redshift space distortions as $(1+f\mu^2)^2$ \citep{Kaiser:1987qv}, where $f$ the growth rate of structure and $\mu$ the cosine of the angle to the line-of-sight. When spherically averaged to compute the matter power spectrum monopole, $P_{\delta \delta}(k)$, this RSD factor gives an amplitude boost of $1.7$ for our fiducial cosmology.   
    \item We then construct an empirical cross-power spectrum model $P_{\hinospace,{\rm g}}$ given by \citep{Masui:2012zc}: 
\begin{equation}
    P_{\hinospace,{\rm g}}(k) = T_\hinospace b_\hinospace b_{\rm g}r_{\hinospace,{\rm opt}} P_{\delta \delta}(k) \, .
    \label{eq:model}
\end{equation}
The model is run through the same pipeline as the data to include weighting, beam\footnote{The telescope beam is modelled as a Gaussian with transverse smoothing scale $R$. This is related to the beam angular resolution, $\theta_{\rm FWHM}$, by
$R=\chi(z)\theta_{\rm FWHM}/(2\sqrt{2\mathrm{ln}2})$,
with $\chi(z)$ being the radial comoving distance to redshift $z$. In cross-correlation, the beam induces a smoothing in the transverse direction as ${\rm e}^{-k^2R^2(1-\mu^2)/2}$.}, and window function effects, as described in \citet{Wolz:2015lwa}. We will comment further on our modelling choices at the end of this section.
\item We fit the unknown prefactor $\Omega_\hinospace b_\hinospace r_{\hinospace,{\rm opt}}$ to the data. We perform fits for all three ranges of scales (Cases I, II, and III in \autoref{tab:constraints}). We find a good reduced chi-squared $\chi_{\rm red}^2 \sim 1$ for our choice of model in all cases and samples.
We also note that excluding the measurements at $k<0.08 \, h{\rm Mpc}^{-1}$ (where there are too few modes in the volume) does not make a discernible difference to our results.
\item We report our $\Omega_\hinospace b_\hinospace r_{\hinospace,{\rm opt}}$ at three different effective scales $k_{\rm eff}$, which are estimated by weighting each $k$-point in the cross-power by its $(\mathrm{S_{\rm best-fit}/N})^2$, for Cases I, II, and III. As we already mentioned, we do this because most of our measurements lie at the nonlinear regime. Assigning an effective scale also allows for a better interpretation of the implications for the values of $r_{\hinospace,{\rm opt}}$.
\end{itemize}

\begin{table*}
    \caption{Best-fit and $1\sigma$ statistical errors on $10^3\Omega_\hinospace b_\hinospace r_{\hinospace,{\rm opt}}$ at a mean redshift $z\simeq 0.8$ for $N_{\rm IC}=20,36$, together with the effective scale $k_{\rm eff}$, detection significance, and reduced chi-squared $\chi^2_{\rm red} = \chi^2 / {\rm dof}$} for $N_{\rm IC}=36$ (Cases I, II, and III; see main text for details).\\
    \label{tab:constraints}
    \centering
    \begin{tabular}{lccccc}
    & \bf{GBT$\times$WiggleZ} & \bf{GBT$\times$ELGs} & \bf{GBT$\times$LRGs} & $k_{\rm eff} [h/{\rm Mpc}]$\\
    \hline
    {\bf{Case I} [$k < 0.8 \, h/{\rm Mpc}$]} & & & &\\
    {\bf{NIC=20}:} & $0.35 \pm 0.09$ & $0.20 \pm 0.06$  & $0.12 \pm 0.06$  & - \\
    {\bf{NIC=36}:} & $0.38 \pm 0.08$ ($4.4\sigma$, $\chi^2_{\rm red} \simeq 16/18$)  & $0.26 \pm 0.06$ ($4.5\sigma$, $22.6/18$) & $0.16 \pm 0.06$ ($2.9\sigma$, $22.9/18$) & 0.48 \\
    \hline
    %%%%%%%%%%%%%%%%%%%%%%
    %%%%%%%%%%%%%%%%%%%%%%
    {\bf{Case II} [$k < 0.45 \, h/{\rm Mpc}$]} & & & &\\
    {\bf{NIC=20}:} & $0.53 \pm 0.12$ & $0.36 \pm 0.09$  & $0.28 \pm 0.09$  & - \\
    {\bf{NIC=36}:} & $0.58 \pm 0.09$ ($4.8\sigma$, $\chi^2_{\rm red} \simeq 8.3/14$)  & $0.40 \pm 0.09$ ($4.9\sigma$, $16/14$) & $0.35 \pm 0.08$ ($4.4\sigma$, $12.3/14$) & 0.31 \\
    \hline

    %%%%%%%%%%%%%%%%%%%%%%
    %%%%%%%%%%%%%%%%%%%%%%
    {\bf{Case III} [$k < 0.35 \, h/{\rm Mpc}$]} & &  & & \\
    {\bf{NIC=20}:} & $0.58 \pm 0.17$ & $0.48 \pm 0.12$  & $0.38 \pm 0.12$  & - \\
    {\bf{NIC=36}:} & $0.70 \pm 0.12$ ($4.4\sigma$, $\chi^2_{\rm red} \simeq 6.7/12$)  & $0.55 \pm 0.11$ ($5\sigma$, $11.6/12$) & $0.45 \pm 0.10$ ($4.2\sigma$, $10/12$) & 0.24 \\
    \end{tabular}
\end{table*}

\begin{figure}
    \centering
    \includegraphics[width=\columnwidth]{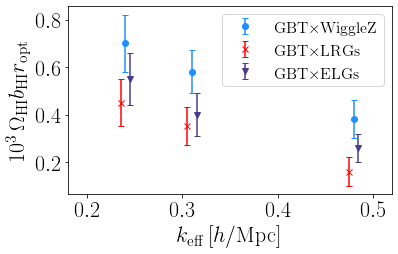}
    \caption{Best-fit and $1\sigma$ statistical errors on $10^3\Omega_\hinospace b_\hinospace r_{\hinospace,{\rm opt}}$ at a mean redshift $z\simeq 0.8$ for $N_{\rm IC}=36$, together with the effective scale $k_{\rm eff}$ (staggered for illustration purposes).}
    \label{fig:omHI_bHI_ropt_constraints}
\end{figure}

Our derived constraints are shown in \autoref{tab:constraints}, for $N_{\rm IC}=20$ and $N_{\rm IC}=36$ (for the smaller $N_{\rm IC}$ cases the errors are too large due to residual foreground variance). In the GBT-WiggleZ Case I, we find excellent agreement with the \citet{Masui:2012zc} results for the 1hr field, $10^3\Omega_\hinospace b_\hinospace r_{\hinospace,{\rm Wig}} = 0.34 \pm 0.07$. Using this case as our benchmark, the lower result in the GBT-ELGs case implies a smaller correlation coefficient between these galaxies and \hinospace, and even smaller in the GBT-LRGs case. The results imply that red galaxies are much more weakly correlated with \hi on the scales we are considering, suggesting that \hi is more associated with blue star-forming galaxies and tends to avoid red galaxies. The same trend is followed in the restricted ranges of scales Cases II and III, albeit with different derived best-fit amplitudes. 
This is in qualitative agreement with what was found in \cite{Anderson_2018} when separating the 2dF survey sample into red and blue galaxies, albeit at a much lower redshift $z=0.08$. The effective scales of the three Cases are different: Case I has 
$k_{\rm eff} = 0.48 \, \, h/{\rm Mpc}$, Case II has $k_{\rm eff} = 0.31 \, \, h/{\rm Mpc}$, and Case III has $k_{\rm eff} = 0.24 \, \, h/{\rm Mpc}$. The different derived best-fit amplitudes are within expectation as $r_{\hinospace,{\rm opt}}$ and $b_\hinospace$ are predicted to be scale-dependent. Therefore, we also expect that if another survey targets larger (linear) scales, e.g. $k < 0.1 \, h/{\rm Mpc}$, it will derive different $\Omega_\hinospace b_\hinospace r_{\hinospace,{\rm opt}}$. 
%%%%
To illustrate the variation between cases, we also present the $N_{\rm IC}=36$ results in \autoref{fig:omHI_bHI_ropt_constraints}.
%%%%

\begin{figure}
    \centering
    \includegraphics[width=\columnwidth]{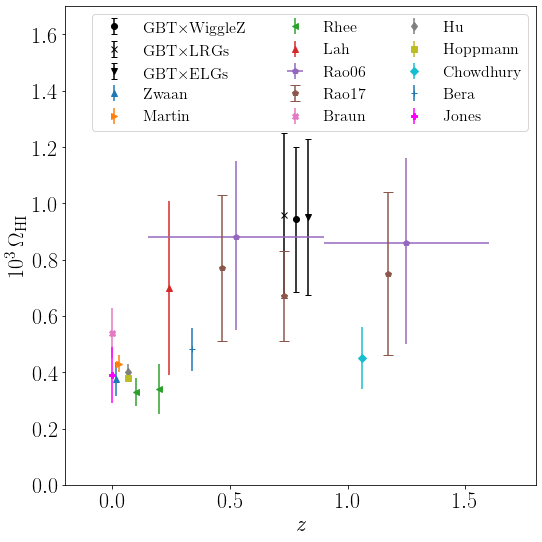}
    \caption{Estimates for $\Omega_\hinospace$ from this work compared to other measurements in the literature. All our estimates are at the central redshift $z=0.78$ but they have been staggered for illustration purposes. We used the results from \autoref{tab:constraints} Case III ($k_{\rm eff}=0.24 \, h/{\rm Mpc}$) for deriving these estimates. \citet{Masui:2012zc} estimated $10^3\Omega_{\hinospace}$ between $0.45$ and $0.75$}. Our assumptions and methodology are detailed in the main text. 
    \label{fig:omHI_constraints}
\end{figure}

We can proceed with the interpretation of our results making some further assumptions. First of all, since the correlation coefficient $r<1$, our results put a lower limit on $\Omega_\hinospace b_\hinospace$. It would also be interesting to attempt to determine $\Omega_\hinospace$ from our measurements taking some external estimates for $b_\hinospace$ and $r_{\hinospace,{\rm opt}}$. The linear bias of \hi is expected to be $\sim 0.65$ to $\sim 1$ at these redshifts \citep{Marin:2009aw}, and we will assume $r_{\rm \hinospace, Wig} = 0.9$ \citep{Khandai:2010hs}. Using our simulations (taking their ratios at $k_{\rm eff}$ for Case III, which is the case where non-linearities are expected to be milder), we can estimate $r_{\rm \hinospace, ELG} \sim 0.7$ and $r_{\rm \hinospace, LRG} \sim 0.6$. Combining these values with the results in \autoref{tab:constraints} and our assumption of perfect knowledge of the galaxy samples biases, we get the $\Omega_\hinospace$ estimates shown in \autoref{fig:omHI_constraints}. These are shown together with other available constraints from the literature \citep{Braun_2012,Zwaan:2005cz,Rao:2005ab,Lah:2007nk,Martin:2010ij,Rhee:2013fma,hoppmann2015blind, Rao2017, Jones2018, Bera:2019gtq, Hu:2019xmd, Chowdhury:2020uqa}. For recent compilations of $\Omega_\hinospace$ measurements in the redshift range $0<z<5$, see \citet{Crighton:2015pza, Neeleman2016, Hu:2019xmd} .

As a final note, we caution the reader that these estimates are crude given the number of assumptions we have made. In principle, the degeneracy between $\Omega_\hinospace$ and $b_\hinospace$ can be broken with the use of redshift space distortions \citep{Wyithe:2008th,Masui:2012zc}, but we need higher quality \hi intensity mapping data with a much better signal-to-noise ratio to achieve this \citep{2010PhRvD..81j3527M,Pourtsidou:2016dzn}. We also stress that while our empirical model (\autoref{eq:model}) has provided an acceptable statistical fit to our data sets, it is not appropriate for high-precision future data. Following what is done in optical galaxy surveys (see e.g. \citet{blake2011,Beutler_2014}), with better data we would need to use more sophisticated models and perform a comprehensive \hi power spectrum multipole expansion analysis \citep{Cunnington:2020mnn}. For example, for the cross-correlation case a more appropriate model to use would be:
\begin{equation}
P_{\hinospace,g}(k,\mu) = T_{\hinospace}b_gb_{\hinospace}
\frac{[r_{\hinospace,{\rm opt}}+(\beta_{\hinospace}+\beta_g)\mu^2+\beta_{\hinospace}\beta_g\mu^4]}{1+(k\mu\sigma_v/H_0)^2}P_{\rm m}(k) \, ,
\end{equation}
with $\beta_i = f/b_i$ and $\sigma_v$ the velocity dispersion parameter. Further, to appropriately model the power spectrum at scales above $k\sim 0.15 \, h{\rm Mpc}^{-1}$ at $z\sim 1$ we would also need to account for scale-dependent bias and $r_{\hinospace,{\rm opt}}$, and construct perturbation theory based models \citep{Villaescusa-Navarro:2018vsg,Castorina:2019zho} including observational effects \citep{Blake:2019ddd,Soares:2020zaq}. To summarise, with our currently available measurements we are very constrained in the number of parameters we can simultaneously fit, and we cannot break any degeneracies unless we use several assumptions and external estimates, hence our empirical choice of model. Furthermore, for precision cosmology studies with future data we will need to take into account the cosmology dependence of the transfer function \citep{Soares:2020zaq}.

\section{Conclusions}
\label{sec:conclusions}
In this work, we performed the first ever comparison of the \hi intensity mapping detections in cross-correlation with multiple galaxy surveys. We use an extended version of the previously published GBT \hi intensity mapping data located in the 1hr field in combination with the WiggleZ Dark Energy Galaxy survey, and the SDSS eBOSS ELG and LRG samples. 

For the GBT data, we subtract the foregrounds and mitigate some systematics via \textsc{FastICA} for $N_{\rm IC} \in \{4,8,20,36\}$. In addition, for the first time for \textsc{FastICA}, we construct a transfer function for the \hi signal loss via mock simulations. We find that there can be a high signal loss up to $50\%$ for $k<0.2 \, h{\rm Mpc}^{-1}$, as foreground removal affects the line-of-sight modes on these scales for all $N_{\rm IC}$. The transfer function converges towards unity for smaller scales, however, for $N_{\rm IC}=36$, we find there is a minimum of $20\%$ signal loss on all scales. The amplitude of the transfer function varies between seasons, indicating that the systematics strongly affect the \hi signal loss. 

For the \hi intensity mapping auto-power spectrum, we find that the amplitude of the  cross-season power spectrum converges for increasing number of ICs. The amplitude is in agreement with previous work in \cite{Masui:2012zc, Switzer_2013, Wolz:2015lwa}, and should be interpreted as an upper limit for detection. 

We investigate the shapes of the galaxy cross-power spectrum, particularly, the correlation between the WiggleZ and the LRG data. We observe a drop in amplitude on the small scales $k\approx0.8$ for the LRG-Wigglez correlation, which can be assumed as a proxy for the \hinospace-LRG correlation, as WiggleZ galaxies are assumed to be \hi-rich and hence a similar tracer to \hi intensity maps. 
We find that the amplitudes of the \hinospace-galaxy cross-correlations do not strongly depend on the $N_{\rm IC}$ of our foreground subtraction. We find a significant drop in amplitude in the \hinospace-LRG correlation at large scales, in agreement with previous findings in \cite{Anderson_2018}.

We construct a mock data set including \hi information and optical galaxy magnitudes based on the outputs of the semi-analytic model DARKSAGE and qualitatively compare the results to our data. Our mock catalogues predict the WiggleZ sample to contain the \hinospace-richest galaxies. Due to the selection of bright objects, the LRG sample also has relatively \hinospace-rich objects, and the averaged mass is in a similar range as the ELG sample. The simulations confirm a drop in amplitude in the LRG-WiggleZ correlation, but not in the \hinospace-LRG correlation. This could be due to failure of our simulation (not matching selection of our galaxies), or the decrease in amplitude caused by observational effects. The present signal-to-noise ratio is not high enough to investigate this further. 

Finally, we use the cross-correlation measurements to constrain the quantity $\Omega_\hinospace b_\hinospace r_{\hinospace,{\rm opt}}$, where $\Omega_\hinospace$ is the \hi density fraction, $b_\hinospace$ is the \hi bias, and $r_{\hinospace,{\rm opt}}$ the galaxy-hydrogen correlation coefficient. 
We consider three different ranges of scales, which correspond to three different effective scales $k_{\rm eff}$ for our derived constraints. 
At $k_{\rm eff}=0.31 \, h/{\rm Mpc}$
we find $\Omega_{\textrm{H\textsc{i}}} b_{\textrm{H\textsc{i}}} r_{\textrm{H\textsc{i}},{\rm Wig}} = [0.58 \pm 0.09 \, {\rm (stat) \pm 0.05 \, {\rm (sys)}}] \times 10^{-3}$ for GBT-WiggleZ, $\Omega_{\textrm{H\textsc{i}}} b_{\textrm{H\textsc{i}}} r_{\textrm{H\textsc{i}},{\rm ELG}} = [0.40 \pm 0.09 \, {\rm (stat) \pm 0.04 \, {\rm (sys)}}] \times 10^{-3}$ for GBT-ELG, and $\Omega_{\textrm{H\textsc{i}}} b_{\textrm{H\textsc{i}}} r_{\textrm{H\textsc{i}},{\rm LRG}} = [0.35 \pm 0.08 \, {\rm (stat) \pm 0.03 \, {\rm (sys)}}] \times 10^{-3}$ for GBT-LRG, at $z\simeq 0.8$. We also report results at $k_{\rm eff}=0.24 \, h/{\rm Mpc}$ and $k_{\rm eff}=0.48 \, h/{\rm Mpc}$.
The  best-fit amplitudes and $1\sigma$ statistical errors for all these cases are shown in \autoref{tab:constraints}.
Our results are amongst the most precise constraints on neutral hydrogen density fluctuations in a relatively unexplored redshift range, using three different galaxy samples. 

Our findings as well as our developed simulations and data analysis pipelines will be useful for the analysis of forthcoming \hi intensity mapping data, and for the preparation of future surveys. 

\section*{Acknowledgements}

We are grateful to Chris Blake for very useful discussions and feedback. We thank the anonymous referee for their insightful questions and helpful suggestions. A.P. is a UK Research and Innovation Future Leaders Fellow [grant number MR/S016066/1], and also acknowledges support by STFC grant ST/S000437/1.

T.C.C. acknowledges support by the JPL Research and Technology Development Fund. Part of the research described in this paper was carried out at the Jet Propulsion Laboratory, California Institute of Technology, under a contract with the National Aeronautics and Space Administration. 
S.A. is supported
by the MICUES project, funded by the EU H2020 Marie Skłodowska-Curie
Actions grant agreement no. 713366 (InterTalentum UAM).
U.-L.P. receives support from Natural Sciences and Engineering Research Council of Canada (NSERC) [funding reference number RGPIN-2019-067, 523638-201], Canadian Institute for Advanced Research (CIFAR), Canadian Foundation for Innovation (CFI), Simons Foundation, and Alexander von Humboldt Foundation. S.C. acknowledges support by STFC grant ST/S000437/1. G.R. acknowledges support from the National Research Foundation of Korea (NRF) through Grant No. 2020R1A2C1005655 funded by the Korean Ministry of Education, Science and Technology (MoEST).
Funding for the Sloan Digital Sky 
Survey IV has been provided by the 
Alfred P. Sloan Foundation, the U.S. 
Department of Energy Office of 
Science, and the Participating 
Institutions. 

SDSS-IV acknowledges support and 
resources from the Center for High 
Performance Computing  at the 
University of Utah. The SDSS 
website is www.sdss.org.

SDSS-IV is managed by the 
Astrophysical Research Consortium 
for the Participating Institutions 
of the SDSS Collaboration including 
the Brazilian Participation Group, 
the Carnegie Institution for Science, 
Carnegie Mellon University, Center for 
Astrophysics | Harvard \& 
Smithsonian, the Chilean Participation 
Group, the French Participation Group, 
Instituto de Astrof\'isica de 
Canarias, The Johns Hopkins 
University, Kavli Institute for the 
Physics and Mathematics of the 
Universe (IPMU) / University of 
Tokyo, the Korean Participation Group, 
Lawrence Berkeley National Laboratory, 
Leibniz Institut f\"ur Astrophysik 
Potsdam (AIP),  Max-Planck-Institut 
f\"ur Astronomie (MPIA Heidelberg), 
Max-Planck-Institut f\"ur 
Astrophysik (MPA Garching), 
Max-Planck-Institut f\"ur 
Extraterrestrische Physik (MPE), 
National Astronomical Observatories of 
China, New Mexico State University, 
New York University, University of 
Notre Dame, Observat\'ario 
Nacional / MCTI, The Ohio State 
University, Pennsylvania State 
University, Shanghai 
Astronomical Observatory, United 
Kingdom Participation Group, 
Universidad Nacional Aut\'onoma 
de M\'exico, University of Arizona, 
University of Colorado Boulder, 
University of Oxford, University of 
Portsmouth, University of Utah, 
University of Virginia, University 
of Washington, University of 
Wisconsin, Vanderbilt University, 
and Yale University.
Simulation data used in this work was generated using Swinburne University's Theoretical Astrophysical Observatory (TAO) and is freely accessible at https://tao.asvo.org.au/. The DARK SAGE semi-analytic galaxy formation model is a public codebase available for download at https://github.com/arhstevens/DarkSage. The Millennium Simulation was carried out by the Virgo Supercomputing Consortium at the Computing Centre of the Max Plank Society in Garching, accessible at http://www.mpa-garching.mpg.de/Millennium/.
We acknowledge the use of open source software \citep{scipy:2001,Hunter:2007,  mckinney-proc-scipy-2010, numpy:2011}.

\emph{Author contributions}: L.W. and A.P. conceived the idea, designed the methodology, led the data analysis, and drafted the paper. All authors contributed to the development and writing of the paper, or made a significant contribution to the data products.

\section*{Data Availability}

The raw GBT intensity mapping data (the observed time stream data) is publicly available according to the NRAO data policy, which can be found at \url{https://science.nrao.edu/observing/proposal-types/datapolicies}. The data products, such as maps and foreground removed maps, will be shared on reasonable request to the corresponding author. We foresee a public release of the GBT data products once the analysis of the maps is finalised and the results are published in scientific journals. The SDSS-IV DR16 data is available at \url{https://www.sdss.org/dr16/}.

The DR16 LSS catalogues are publicly available: \url{https://data.sdss.org/sas/dr16/eboss/lss/catalogs/DR16/}.

%%%%%%%%%%%%%%%%%%%%%%%%%%%%%%%%%%%%%%%%%%%%%%%%%%


\bibliographystyle{mnras}
\bibliography{main_bib}

%%%%%%%%%%%%%%%%%%%%%%%%%%%%%%%%%%%%%%%%%%%%%%%%%%

%%%%%%%%%%%%%%%%% APPENDICES %%%%%%%%%%%%%%%%%%%%%

\appendix
\section{Sample selection for optical mock galaxies}
\label{app:samples}

For our sample selection in the simulation, we use the selection which includes the magnitude limits of the observations as well as the target selection. 

For WiggleZ, we use the selection cuts outlines in \cite{Drinkwater}, as we have previously done in \cite{Wolz:2015lwa}. The selection is based on the GALEX UV filters NUV and FUV, as well as the SDSS $r$ filter, as follows.
\begin{equation}
\begin{split}
    {\rm NUV} &< 22.8 \\
    20 &< r < 22  \\
    -0.5 & < ({\rm NUV} - r) < 2.
\end{split}
\end{equation}
For eBOSS ELG, we follow 
\begin{equation}
\begin{split}
   21.825< &g < 22.825 \\
 (-0.068(r-z)+0.457) < &(g-r) < (0.112(r-z)+0.773) \\
 (0.218(g-r)+0.571) < &(r-z) < (-0.555(g-r)+1.901).
\end{split}
\end{equation}

For eBOSS LRG, we follow \cite{}, where we use the infra red filter IRAC1 as a close approximation for the WISE filter. 
\begin{equation}
\begin{split}
(19.9<& i <21.8) \\
(&z < 19.95) \\
(& {\rm IRAC1} <20.299)\\
& (r - i) > 0.98)\\
&(r - {\rm IRAC1})> 2(r-i)).
\end{split}
\end{equation}

\section{Covariance and Error estimates}
\label{appb}
Here we show the covariances and error estimates as described in \autoref{analysistests}.

\begin{figure}
    \centering
    \includegraphics[width = 0.5\columnwidth]{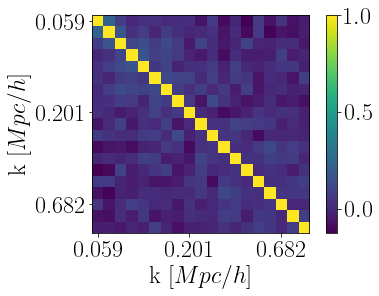}\includegraphics[width = 0.5\columnwidth]{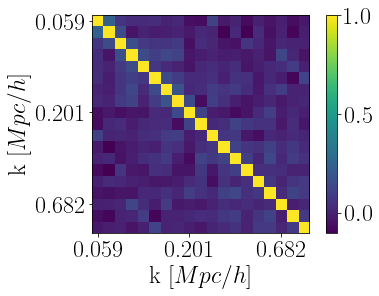}\\
    \includegraphics[width = 0.5\columnwidth]{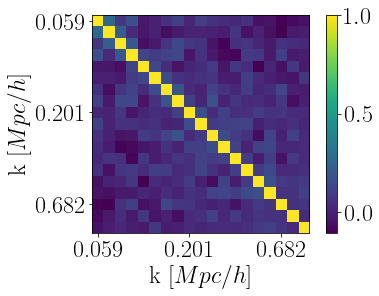}\includegraphics[width = 0.5\columnwidth]{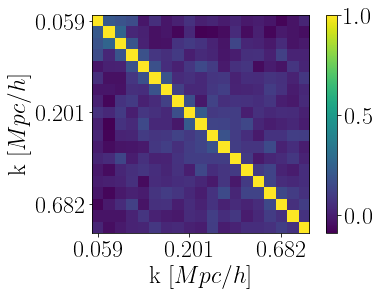}
    \caption{The covariance matrix computed from the power spectrum of the foreground removed lognormal realisations with the original lognormals, as described in \autoref{subsec:transfer} with clockwise increasing numbers of ICs $N_{\rm IC}$. \textit{Upper left panel}: $N_{\rm IC} = 4$; \textit{Upper right panel}: $N_{\rm IC} = 8$; \textit{Lower left panel}: $N_{\rm IC} = 20$; \textit{Lower right panel}: $N_{\rm IC} = 36$. For illustrative purposes the diagonals of the covariance matrices have been normalised to unity; i.e. the correlation matrix is pictured.  }
    \label{fig:covariancelognormals}
\end{figure}

In \autoref{fig:covariancelognormals} we show the covariance based on the cross-power spectra of the 100 lognormal realisations (after injected into the GBT data and cleaned with fastICA) with the original lognormal realisations, see \autoref{subsec:transfer} and \autoref{analysistests} for details. We can see that the application of fastICA does not introduce any significant correlations between $k$-modes or off-diagonal elements. 

\begin{figure}
    \centering
    \includegraphics[width = 0.5\columnwidth]{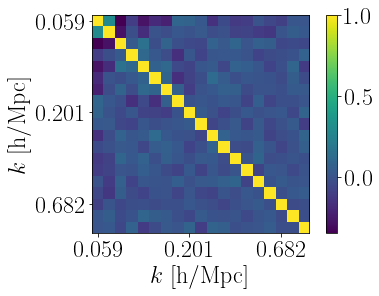}\includegraphics[width = 0.5\columnwidth]{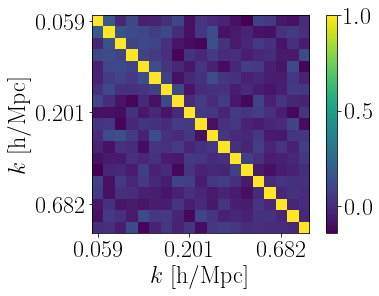}\\
    \includegraphics[width = 0.5\columnwidth]{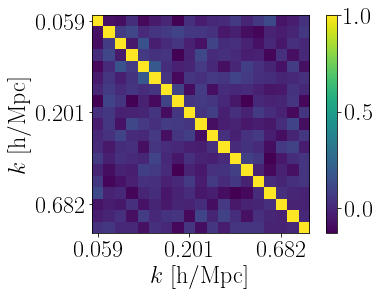}\includegraphics[width = 0.5\columnwidth]{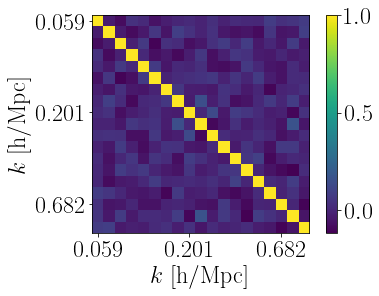}
    \caption{The covariance matrix computed from the cross-correlation of the foreground removed GBT data with WiggleZ random catalogues, as described in \autoref{analysistests} with clockwise increasing numbers of ICs $N_{\rm IC}$. \textit{Upper left panel}: $N_{\rm IC} = 4$; \textit{Upper right panel}: $N_{\rm IC} = 8$; \textit{Lower left panel}: $N_{\rm IC} = 20$; \textit{Lower right panel}: $N_{\rm IC} = 36$. For illustrative purposes the diagonals of the covariance matrices have been normalised to unity; i.e. the correlation matrix is pictured.  }
    \label{fig:covariancewiggleZ}
\end{figure}

In \autoref{fig:covariancewiggleZ}, we show the covariance based on the GBT data with the 100 WiggleZ random catalogues as described in \autoref{subsec:transfer}. We can see that for lower number of ICs, there are non-negligible off-diagonal elements for small $k$-modes, and particularly for $N_{\rm IC}=4$, the amplitude of the off-diagonal elements is increased. Note that the covariance in this work includes both cosmic variance as well as variance from the noise as we average over 100 realisations as well as over the 4 independent GBT data sections.

\begin{figure}
    \centering
    \includegraphics[width=\columnwidth]{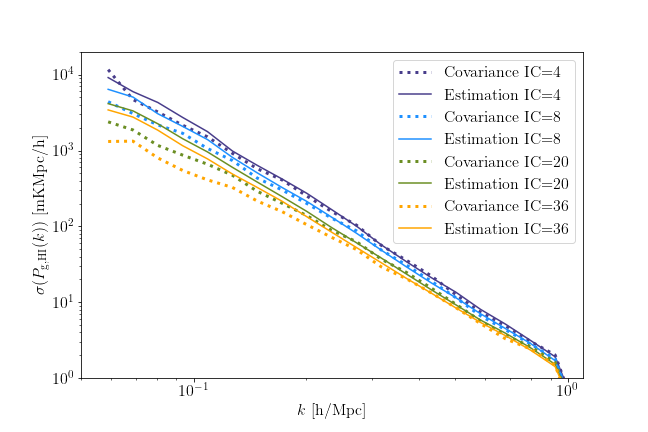}
    \caption{The comparison of the error bars coming from the estimate in \autoref{eq:sgHI} versus the estimate from the covariance matrix of the GBT with the WiggleZ random catalogues.}
    \label{fig:errorcomp}
\end{figure}

In \autoref{fig:errorcomp}, we show the comparison of the errorbars resulting from \autoref{eq:sgHI} and the diagonal of the covariance matrix based on the GBT data with the WiggleZ randoms as shown in \autoref{fig:covariancewiggleZ}. We can see that for small $k$-modes, the error estimate based on the errors from the estimated auto-powerspectra in \autoref{eq:sgHI} is higher than the covariance-based estimate, and both estimates converge towards smaller scales.


%If you want to present additional material which would interrupt the flow of the main paper,it can be placed in an Appendix which appears after the list of references.

%%%%%%%%%%%%%%%%%%%%%%%%%%%%%%%%%%%%%%%%%%%%%%%%%%

% Don't change these lines
\bsp	% typesetting comment
\label{lastpage}
\end{document}